\documentclass[a4paper,12pt,fleqn]{article}

\usepackage{changepage}
\usepackage{amssymb}
\usepackage{slashed}
\usepackage{pifont}
\usepackage[margin=1.0in]{geometry}
\usepackage{fleqn}
\usepackage{amsmath}
\usepackage{amsbsy}
\usepackage{mathrsfs}
\usepackage{textgreek}
\usepackage{hyperref}
\allowdisplaybreaks

\newcommand\psibar{\overline{\psi}}
\newcommand\chibar{\overline{\chi}}

\newcommand\rmi{\mathrm{i}}
\newcommand\rmc{\mathrm{c}}

\newcommand\rmint{\mathrm{int}}
\newcommand\rmMD{\mathrm{MD}}

\newcommand\rmd{\mathrm{d}}

\newcommand\rmD{\mathrm{D}}
\newcommand\rmH{\mathrm{H}}
\newcommand\rmM{\mathrm{M}}

\newcommand\rmem{\mathrm{em}}
\newcommand\rmRe{\mathrm{Re}}
\newcommand\rmIm{\mathrm{Im}}

\newcommand\scF{\mathcal{F}}
\newcommand\sL{\mathscr{L}}

\newcommand\tT{\widetilde{T}}

\newcommand\tP{\widetilde{P}}

\newcommand\sdual{{}^{*}\!s}

\begin{document}




\section*{{\LARGE Maxwell-Dirac stress-energy tensor in terms of Fierz bilinear currents}}
\vspace{0.5cm}




\begin{adjustwidth}{1cm}{}
{\large \textbf{S M Inglis and P D Jarvis}}

\vspace{0.2cm}
{\small\noindent School of Mathematics and Physics, University of Tasmania, Sandy Bay Campus, Private Bag 37, Hobart, Tasmania, 700}

\vspace{0.2cm}
{\small\noindent Email: {\tt sminglis@utas.edu.au, Peter.Jarvis@utas.edu.au}}

\vspace{0.5cm}
{\small
\noindent\textbf{Abstract.} We analyze the stress-energy tensor for the self-coupled Maxwell-Dirac system in the bilinear current formalism, using two independent approaches. The first method used is that attributed to Belinfante: starting from the spinor form of the action, the well-known canonical stress-energy tensor is augmented, by extending the Noether symmetry current to include contributions from the Lorentz group, to a manifestly symmetric form. This form admits a transcription to bilinear current form. The second method used is the variational derivation based on the covariant coupling to general relativity. The starting point here at the outset is the transcription of the action using, as independent field variables, both the bilinear currents, together with a gauge invariant vector field (a proxy for the electromagnetic vector potential). A central feature of the two constructions is that they both involve the mapping of the Dirac contribution to the stress-energy from the spinor fields to the equivalent set of bilinear tensor currents, through the use of appropriate Fierz identities. Although this mapping is done at quite different stages, nonetheless we find that the two forms of the bilinear stress-energy tensor agree. Finally, as an application, we consider the reduction of the obtained stress-energy tensor in bilinear form, under the assumption of spherical symmetry.}
\end{adjustwidth}







\section{Introduction}\label{Introduction}

In classical field theory, it is well known that a conserved energy-momentum tensor for a physical system can be constructed by the Noether method, exploiting invariance of the Lagrangian density under space-time translations. However, the derived symmetry current using this method, the so-called ``canonical'' form \cite{Greiner1990}, has the unfortunate drawbacks of being neither symmetric in its two tensor indices, nor gauge invariant.

Many attempts have been made to rectify these problems, two of the most prominent being the Belinfante form \cite{Belinfante1939}, and the variational form from general relativity \cite{Wald1984}. The basis of the Belinfante approach is to extend the invariance of the Lagrangian to include contributions from Lorentz transformations, so that the Noether symmetry current becomes that associated with the full Poincar\'{e} group. In this way, the canonical term is symmetrized \cite{Weinberg1995}, and extra ``correction'' terms are present, which are attributed to the spin contribution to the stress-energy. Under the assumption of a torsion-free space-time the asymmetric spin terms vanish, but in general, without this assumption an asymmetric stress-energy tensor is valid \cite{Hehl1976}.

The variational approach uses the action principle to relate the variation of the Hilbert action of space-time to that of matter, then invoking Einstein's equations to identify the matter part to the stress-energy tensor. The result is an expression for the stress-energy which is proportional to the functional derivative of the action of matter, with respect to the inverse metric. The presence of the metric ensures that the stress-energy tensor is manifestly symmetric.

Regardless of the independent nature of their derivations, Goedecke pointed out \cite{Goedecke1974} that in the limit of flat space-time, the Belinfante and variational forms of the stress-energy tensor must agree. The equivalence in the integral spin field case was proven by Rosenfeld \cite{Rosenfeld1940}, and Goedecke provided evidence for equivalence in the half-integral spin field case via a series of examples, but was not able to provide a general proof. Such a proof for the half-integral case was published shortly afterwards by Lord \cite{Lord1976}, using the vierbein formalism.

Much effort has been made to ``improve'' the stress-energy tensor, either by generalizing it beyond the Belinfante/variational forms, or by altering it so that it is compatible with a given theory. An example of the generalization aspect is the work done by Gotay and Marsden, who model the stress-energy in terms of fluxes of the multimomentum map across spacetime hypersurfaces \cite{GotayMarsden1992}. The Gotay-Marsden stress-energy tensor naturally includes the spin ``correction terms'' present in the Belinfante formula, but in a more generalized fashion, as well as coinciding with the variational form in the presence of a space-time metric. Work has also been done by Callan, Coleman and Jackiw on making the cut-off dependent symmetric stress-energy tensor compatible with renormalized perturbation theory, by constructing appropriate counter terms in order to make it finite at arbitrarily large cut-off values \cite{CallanColemanJackiw1970}. The renormalization compatible stress-energy tensor is also compatible with an altered, but phenomenologically consistent, version of general relativity, and simplifies the currents associated with scale and conformal transformations in which the stress-energy appears.

The work we present in this paper can be viewed as analogous to one of Goedecke's examples, namely the coupled Maxwell-Dirac fields, but in an alternate formalism where the spinor fields are mapped bilinearly to a set of tensor fields. The motivation for working in such a formalism is that it removes any dependence of the physical states on gauge transformations, thereby constituting a description of the physics in terms of observable densities only. Furthermore, eliminating gauge dependence simplifies the process of reducing the Maxwell-Dirac equations under geometrical symmetries of the Poincar\'{e} group, since only the calculation of the generic subgroup symmetric forms of the tensor fields need to considered. Questions of gauge fixing and symmetry reductions of gauge dependent fields are removed entirely. Some examples of symmetry reductions of the Maxwell-Dirac equations in the gauge invariant bilinear tensor formalism were presented in \cite{InglisJarvis2014}.

The background to this approach is as follows. Interrelationships between these ``bilinears'' can be derived \cite{Fierz1937} by taking their quadratic products, and expanding out the $\chi\psibar$ spinor matrices in terms of the Dirac-Clifford algebra basis of $4\times4$ matrices. For the $\chi\equiv\psi$ case, there are sixteen real, gauge invariant bilinears, equal to the number of basis elements. Comparing this number with the eight real components of the Dirac spinors, and taking the gauge invariance into account, implies that there must be nine independent algebraic equations \cite{Kaempffer1981}. Furthermore, the bilinear map is invertible according to the spinor reconstruction theorem by Crawford \cite{Crawford1985}, so given a state described by the set of sixteen bilinear tensor fields, the Dirac spinor field is determined up to a phase. This bilinear reformulation is generalizable to the non-Abelian isospinor representations, at least in the $SU(2)$ doublet case, where an analogous set of Fierz identities involving \emph{sixty four} bilinears can be shown to exist \cite{InglisJarvis2012}.

There are many other identities involving the sixteen bilinears beyond the fundamental set, which can be derived by using the Fierz expansion along with known identities. The extension of this set of sixteen to other classes of bilinears, such as gauge dependent objects where some of the spinors are charge conjugated (i.e., $\chi\equiv\psi^{\rmc}$), or contain derivatives, is a necessary part of the mathematical framework for the description of the Maxwell-Dirac equations in bilinear form \cite{InglisJarvis2014}. A serious attempt at generalizing the fundamental set of nine Fierz identities to include a larger set of bilinear ``currents'' was presented by Takahashi \cite{Takahashi1983}, although bilinears containing spinor derivatives were only discussed briefly.

The bilinear representation of electrodynamics is compatible with the Maxwell-Dirac equations, which model the self-consistent behaviour of an electrically charged fermion field interacting with its own electromagnetic field. The derivation of this system relies on the gauge covariant Dirac equation being invertible, with the gauge field, or vector potential $A_{\mu}$, determined by the state described by the spinors, or equivalently, the bilinears. The inversion in the $U(1)$ (electromagnetic) case was originally considered by Eliezer \cite{Eliezer1958}, however he pointed out that the matrix required for the inversion had vanishing determinant, which discounted the possibility for a unique solution. More recently it was shown \cite{BoothLeggJarvis2001} that if the physical requirement that the vector potential is real is imposed, then the Dirac equation is indeed invertible, provided that a set of consistency conditions are satisfied. Incidentally, a formal inversion of the Dirac equation for the $SU(2)$ Yang-Mills gauge field has also been achieved \cite{InglisJarvis2012}  which extends to $U(2)\cong SU(2)\times U(1)$, but whether or not the inversion can be extended even further to the more physically relevant $U(1)\times SU(2)\times SU(3)$ case is an open question. The set of equations obtained when the inverted Dirac equation is substituted into the Maxwell equations, along with the appropriate consistency conditions and Fierz identities, constitutes the Maxwell-Dirac equations. We shall not discuss the details of this system further here, other than to state the aim of this paper, which is to derive a bilinear form of the stress-energy tensor for this system, and explore how it reduces in the presence of symmetries. For further information on the bilinear current, or ``relativistic hydrodynamical'' formulation of the Maxwell-Dirac system, we direct the reader to \cite{Takabayasi1957}, and \cite{InglisJarvis2014} where symmetry reductions are also studied.

The first topic we undertake in this paper is the derivation of the bilinear form of the Maxwell-Dirac-Belinfante tensor in Section \ref{Maxwell-Dirac stress-energy tensor via Belinfante}. Following a brief derivation of the Belinfante tensor for a free Dirac particle in the spinor representation, we introduce the core concepts of the bilinear mapping and Fierz expansions. These ideas are then applied to the derivative-dependent bilinear terms that appear naturally in the free Dirac-Belinfante tensor, and an appropriate Fierz identity is derived, allowing us to rewrite the explicitly spinor-dependent form of the tensor exclusively in terms of bilinears. Note that the Fierz transcription to bilinears occurs as the \emph{final stage} of the calculation, once the Belinfante tensor in Dirac spinor form has already been derived. A more detailed version of this derivation is relegated to \ref{Derivation of the Belinfante Fierz identity}. The known tensorial forms of the electromagnetic interaction and Maxwell field stress-energies are then added to the free Dirac contribution, resulting in a manifestly symmetric and gauge independent bilinear form of the Maxwell-Dirac-Belinfante tensor.

Section \ref{Maxwell-Dirac stress-energy tensor via general relativity} presents an independent derivation of the Maxwell-Dirac stress-energy tensor, which in this case uses the variational form known from general relativity. Beginning with the Lagrangian density for an electromagnetically interacting Dirac particle, and initially ignoring the Maxwell field contribution since we are mainly interested in the behaviour of the bilinear Dirac contribution, we convert it to its analogous bilinear form, using a contracted form of the Fierz identity obtained in Section \ref{Maxwell-Dirac stress-energy tensor via Belinfante}. This is in stark contrast to the calculational method in Section \ref{Maxwell-Dirac stress-energy tensor via Belinfante}, where the Fierz transcription took place last. A brief review of how the variational stress-energy is obtained is then given. Then, using the general relativistically covariant form of the bilinear Dirac Lagrangian, the variational stress-energy is obtained, and is found to be in agreement with the Maxwell-Dirac-Belinfante tensor.

Finally, in Section \ref{Symmetry reduction of the Maxwell-Dirac stress-energy tensor}, the bilinear Maxwell-Dirac stress-energy tensor is subjected to the restrictions imposed by an example symmetry subgroup of the Poincar\'{e} group, namely the spherical symmetry group $SO(3)$. This extends the discussion of the symmetry reductions of the Maxwell-Dirac equations from \cite{InglisJarvis2014}, essentially providing the mathematical framework required to calculate the mass-energy and momentum fluxes corresponding to solutions obtained under this symmetry and others, in principle. The discussion of such Maxwell-Dirac solutions and their physical properties, as well as the treatment of other select symmetry groups, are intended to be presented in follow-up works.

\section{Maxwell-Dirac stress-energy tensor via Belinfante}\label{Maxwell-Dirac stress-energy tensor via Belinfante}
\subsection{Belinfante tensor for a free Dirac particle}
The Belinfante stress-energy tensor is the fully symmetric analogue of the well-known asymmetric ``canonical'' form, which for free Dirac particles is\footnote{For more details on mathematical conventions, refer to \cite{InglisJarvis2014} and \cite{ItzyksonZuber1980}.}
\begin{equation}\label{Canonical stress-energy tensor}
\tT^{\mu\nu}=\frac{\rmi}{2}\left[\psibar\gamma^{\mu}(\partial^{\nu}\psi)-(\partial^{\nu}\psibar)\gamma^{\mu}\psi\right],
\end{equation}
that satisfies the conservation condition
\begin{equation}
\partial_{\mu}\tT^{\mu\nu}=0.
\end{equation}
In fact, $\tT^{\mu\nu}$ is the Noether symmetry current corresponding to imposing the invariance of the free Dirac Lagrangian
\begin{equation}\label{Free Dirac Lagrangian}
\sL=\frac{\rmi}{2}\left[\psibar\gamma^{\mu}(\partial_{\mu}\psi)-(\partial_{\mu}\psibar)\gamma^{\mu}\psi\right]-m\psibar\psi
\end{equation}
under the translation group. When rotational contributions are included, the asymmetry can be isolated into a term involving the divergence of the spin current $S^{\sigma\mu\nu}=(1/4)\psibar\{\gamma^{\sigma},\sigma^{\mu\nu}\}\psi$, which can be made to vanish in the absence of torsion \cite{Hehl1976}, \cite{Doughty1990}. Since (\ref{Free Dirac Lagrangian}) is invariant under Lorentz transformations, we can use the formula for the Noether symmetry current divergence
\begin{equation}
\partial_{\mu}\left[\frac{\partial\sL}{\partial(\partial_{\mu}\psi)}\delta\psi+\delta\psibar\frac{\partial\sL}{\partial(\partial_{\mu}\psibar)}\right]=0,
\end{equation}
to obtain the Noether current directly. A vanishing, manifestly antisymmetric expression is obtained
\begin{equation}\label{Vanishing antisymmetric Noether form}
\frac{1}{4}\partial_{\sigma}\left(\psibar\{\gamma^{\sigma},\sigma^{\mu\nu}\}\psi\right)+\tT^{\mu\nu}-\tT^{\nu\mu}=0.
\end{equation}
This can be interpreted as an antisymmetric combination of a symmetric tensor \cite{Weinberg1995}, which we call $\Theta^{\mu\nu}$. A tensor form which reduces to the left-hand side of (\ref{Vanishing antisymmetric Noether form}) upon antisymmetrization is
\begin{equation}
\Theta^{\mu\nu}=\tT^{\mu\nu}+\frac{1}{8}\partial_{\sigma}\left[\psibar\{\gamma^{\sigma},\sigma^{\mu\nu}\}\psi-\psibar\{\gamma^{\mu},\sigma^{\sigma\nu}\}\psi-\psibar\{\gamma^{\nu},\sigma^{\sigma\mu}\}\psi\right].
\end{equation}
Using (\ref{Vanishing antisymmetric Noether form}) to replace the left-most anti-commutator bilinear term, results in the manifestly symmetric combination
\begin{equation}
\Theta^{\mu\nu}=\frac{1}{2}(\tT^{\mu\nu}+\tT^{\nu\mu})-\frac{1}{8}\partial_{\sigma}\left[\psibar\{\gamma^{\mu},\sigma^{\sigma\nu}\}\psi+\psibar\{\gamma^{\nu},\sigma^{\sigma\mu}\}\psi\right],
\end{equation}
where the final term involving the divergences of the spin density
\begin{equation}
-\frac{1}{2}\partial_{\sigma}(S^{\mu\sigma\nu}+S^{\nu\sigma\mu})=-\frac{1}{8}\partial_{\sigma}\left[\psibar\{\gamma^{\mu},\sigma^{\sigma\nu}\}\psi+\psibar\{\gamma^{\nu},\sigma^{\sigma\mu}\}\psi\right]=0,
\end{equation}
vanishes when taking into account the identity in the Dirac algebra
\begin{equation}
\{\gamma^{\mu},\sigma^{\sigma\nu}\}=2\epsilon^{\sigma\nu\mu\rho}\gamma_{5}\gamma_{\rho}.
\end{equation}
We therefore obtain the form of the Belinfante stress-energy tensor for a free Dirac particle
\begin{equation}\label{Belinfante stress-energy for free Dirac particle}
\Theta^{\mu\nu}=\frac{1}{2}(\tT^{\mu\nu}+\tT^{\nu\mu})=\frac{\rmi}{4}\left[\psibar\gamma^{\mu}(\partial^{\nu}\psi)-(\partial^{\nu}\psibar)\gamma^{\mu}\psi\right]+\frac{\rmi}{4}\left[\psibar\gamma^{\nu}(\partial^{\mu}\psi)-(\partial^{\mu}\psibar)\gamma^{\nu}\psi\right],
\end{equation}
in agreement with Goedecke \cite{Goedecke1974}. Note that the Belinfante tensor is conserved
\begin{equation}
\partial_{\mu}\Theta^{\mu\nu}=0,
\end{equation}
and is equivalent to the Noether symmetry current of the Poincar\'{e} group in the absence of torsion.

\subsection{Dirac bilinears and Fierz identities}
The bilinear mapping from the spinor fields $\chibar$ and $\psi$ to the corresponding set of tensor fields is of the form $\chibar\Gamma_{R}\psi$, where $\Gamma_{R}=\{I,\gamma^{\mu},\sigma^{\mu\nu},\gamma_{5}\gamma^{\mu},\gamma_{5}\}$ is the set of $R=1,...,16$ basis elements of the Dirac Clifford algebra. When $\chibar=\psibar$, we obtain the set of manifestly $U(1)$ gauge invariant fields
\begin{equation}\label{Real Dirac bilinears}
\psibar\Gamma_{R}\psi=\{\sigma,j^{\mu},s^{\mu\nu},k^{\mu},\omega\},
\end{equation}
which are all real except for $\omega$, which is pure imaginary, as is the additional tensor $\sdual^{\mu\nu}\equiv\psibar\gamma_{5}\sigma^{\mu\nu}\psi$. Note that a common alternate convention is to use the real pseudoscalar field $\varpi\equiv\psibar\rmi\gamma_{5}\psi$. Alternatively, setting $\chibar=\psibar{}^{\rmc}$, we obtain for the $\Gamma_{R}=\gamma^{\mu}$ element, the gauge dependent bilinear vectors
\begin{align}
&m^{\mu}+\rmi n^{\mu}=\psibar{}^{\rmc}\gamma^{\mu}\psi, \\
&m^{\mu}=\rmRe[\psibar{}^{\rmc}\gamma^{\mu}\psi]=\frac{1}{2}(\psibar{}^{\rmc}\gamma^{\mu}\psi+\psibar\gamma^{\mu}\psi^{\rmc}), \\
&n^{\mu}=\rmIm[\psibar{}^{\rmc}\gamma^{\mu}\psi]=\frac{\rmi}{2}(\psibar\gamma^{\mu}\psi{}^{\rmc}-\psibar{}^{\rmc}\gamma^{\mu}\psi).
\end{align}
The matrix product of two spinors can be expanded in the $\Gamma_{R}$ basis, using the Fierz expansion
\begin{equation}\label{Fierz expansion}
\psi\chibar=\sum_{R=1}^{16}a_{R}\Gamma_{R}=\frac{1}{4}(\chibar\psi)I+\frac{1}{4}(\chibar\gamma_{\mu}\psi)\gamma^{\mu}+\frac{1}{8}(\chibar\sigma_{\mu\nu}\psi)\sigma^{\mu\nu}-\frac{1}{4}(\chibar\gamma_{5}\gamma_{\mu}\psi)\gamma_{5}\gamma^{\mu}+\frac{1}{4}(\chibar\gamma_{5}\psi)\gamma_{5},
\end{equation}
where either $\chibar$ or $\psi$ may contain partial derivatives. There exists a rich set of interrelationships between bilinear tensor fields, which can be derived by ``breaking up'' products of bilinears such as $j^{\mu}k^{\nu}\equiv\psibar\gamma^{\mu}(\psi\psibar)\gamma_{5}\gamma^{\nu}\psi$ into sums of different bilinears, by inserting (\ref{Fierz expansion}). This Fierz expansion technique was used extensively in \cite{InglisJarvis2014}, in the context of the reformulation of the self-coupled Maxwell-Dirac equations in terms of bilinears.

\subsection{Belinfante tensor in bilinear form}\label{Belinfante tensor in bilinear form}
Our current objective is to rewrite (\ref{Belinfante stress-energy for free Dirac particle}) in terms of Fierz bilinears, which means we need to derive a Fierz identity that expresses the spinorial object $[\psibar\gamma_{\nu}(\partial_{\mu}\psi)-(\partial_{\mu}\psibar)\gamma_{\nu}\psi]$ in terms of bilinears. Therefore, we are led to search for Fierz expansions in which this term is likely to appear. One example is
\begin{align}\label{Fierz identity for j_nu[...]}
&j_{\nu}[\psibar(\partial_{\mu}\psi)-(\partial_{\mu}\psibar)\psi]=\frac{\rmi}{3}(\partial_{\mu}j^{\sigma})s_{\nu\sigma}-\frac{\rmi}{3}j^{\sigma}(\partial_{\mu}s_{\nu\sigma})+\frac{1}{3}(\partial_{\mu}\omega)k_{\nu}-\frac{1}{3}\omega(\partial_{\mu}k_{\nu}) \nonumber \\
&\qquad +\frac{1}{3}\sigma[\psibar\gamma_{\nu}(\partial_{\mu}\psi)-(\partial_{\mu}\psibar)\gamma_{\nu}\psi]-\frac{\rmi}{3}\sdual_{\nu\sigma}[\psibar\gamma_{5}\gamma^{\sigma}(\partial_{\mu}\psi)-(\partial_{\mu}\psibar)\gamma_{5}\gamma^{\sigma}\psi] \nonumber \\
&\qquad-\frac{\rmi}{3}k^{\sigma}[\psibar\gamma_{5}\sigma_{\nu\sigma}(\partial_{\mu}\psi)-(\partial_{\mu}\psibar)\gamma_{5}\sigma_{\nu\sigma}\psi].
\end{align}
There are at least three other bilinear products whose Fierz expansions produce the desired term, namely $j_{\nu}[\psibar\gamma_{5}(\partial_{\mu}\psi)-(\partial_{\mu}\psibar)\gamma_{5}\psi]$, $k_{\nu}(\partial_{\mu}\sigma)$ and $k_{\nu}(\partial_{\mu}\omega)$. Their respective expanded forms, along with a much more detailed derivation of the Fierz identity, is given in appendix \ref{Derivation of the Belinfante Fierz identity}. Using these four identities, we can combine them to give
\begin{align}\label{Belinfante Fierz identity, with spinor terms}
&[\psibar\gamma_{\nu}(\partial_{\mu}\psi)-(\partial_{\mu}\psibar)\gamma_{\nu}\psi]=(\sigma\omega)^{-1}\left(-\frac{\rmi}{2}(\partial_{\mu}j^{\sigma})(\omega s_{\nu\sigma}+\sigma\sdual_{\nu\sigma})-k_{\nu}[\sigma(\partial_{\mu}\sigma)+\omega(\partial_{\mu}\omega)]\right. \nonumber \\
&\qquad\ \ +\frac{1}{2}(\partial_{\mu}k_{\nu})(\sigma^{2}+\omega^{2})+j_{\nu}\{\omega[\psibar(\partial_{\mu}\psi)-(\partial_{\mu}\psibar)\psi]+\sigma[\psibar\gamma_{5}(\partial_{\mu}\psi)-(\partial_{\mu}\psibar)\gamma_{5}\psi]\} \nonumber \\
&\left.\qquad\ \ +\frac{\rmi}{2}(\sigma s_{\nu\sigma}+\omega\sdual_{\nu\sigma})[\psibar\gamma_{5}\gamma^{\sigma}(\partial_{\mu}\psi)-(\partial_{\mu}\psibar)\gamma_{5}\gamma^{\sigma}\psi]\right),
\end{align}
which obviously still requires some more work. Using the Fierz identities derived in \cite{InglisJarvis2014}
\begin{align}
&[\psibar(\partial_{\mu}\psi)-(\partial_{\mu}\psibar)\psi]=-(\sigma^{2}-\omega^{2})^{-1}[j^{\nu}(\partial_{\mu}k_{\nu})\omega+\rmi m^{\nu}(\partial_{\mu}n_{\nu})\sigma],\label{Anti-product rule on sigma} \\
&[\psibar\gamma_{5}(\partial_{\mu}\psi)-(\partial_{\mu}\psibar)\gamma_{5}\psi]=-(\sigma^{2}-\omega^{2})^{-1}[j^{\nu}(\partial_{\mu}k_{\nu})\sigma+\rmi m^{\nu}(\partial_{\mu}n_{\nu})\omega],\label{Anti-product rule on omega}
\end{align}
we can replace the spinor terms on the second line of (\ref{Belinfante Fierz identity, with spinor terms}) with bilinears, but this still leaves the spinor terms on the third line. After a straightforward, but tedious, set of Fierz manipulations, we obtain the desired identity
\begin{align}
&\frac{\rmi}{2}(\sigma s_{\nu\sigma}+\omega\sdual_{\nu\sigma})[\psibar\gamma_{5}\gamma^{\sigma}(\partial_{\mu}\psi)-(\partial_{\mu}\psibar)\gamma_{5}\gamma^{\sigma}\psi]=-\frac{1}{2}\sigma\omega[\psibar\gamma_{\nu}(\partial_{\mu}\psi)-(\partial_{\mu}\psibar)\gamma_{\nu}\psi] \nonumber \\
&\ \ +\frac{1}{4}j_{\nu}(\sigma^{2}-\omega^{2})^{-1}[j^{\sigma}(\partial_{\mu}k_{\sigma})(\sigma^{2}+\omega^{2})+2\rmi m^{\sigma}(\partial_{\mu}n_{\sigma})\sigma\omega]-\frac{3}{8}(\sigma^{2}+\omega^{2})(\partial_{\mu}k_{\nu}) \nonumber \\
&\ \ +\frac{3}{8}k_{\nu}[\sigma(\partial_{\mu}\sigma)+\omega(\partial_{\mu}\omega)]-\frac{\rmi}{8}(\partial_{\mu}j^{\sigma})(\sigma\sdual_{\nu\sigma}+\omega s_{\nu\sigma})+\frac{\rmi}{8}j^{\sigma}[\sigma(\partial_{\mu}\sdual_{\nu\sigma})+\omega(\partial_{\mu}s_{\nu\sigma})].
\end{align}
Substituting into (\ref{Belinfante Fierz identity, with spinor terms}) and rearranging yields
\begin{align}\label{Belinfante Fierz identity, with rank-2 terms}
&[\psibar\gamma_{\nu}(\partial_{\mu}\psi)-(\partial_{\mu}\psibar)\gamma_{\nu}\psi] \nonumber \\
&\qquad=(\sigma\omega)^{-1}\left\{-\frac{1}{2}j_{\nu}(\sigma^{2}-\omega^{2})^{-1}[j^{\sigma}(\partial_{\mu}k_{\sigma})(\sigma^{2}+\omega^{2})+2\rmi m^{\sigma}(\partial_{\mu}n_{\sigma})\sigma\omega]\right. \nonumber \\
&\qquad\qquad+\frac{1}{12}(\sigma^{2}+\omega^{2})(\partial_{\mu}k_{\nu})-\frac{5}{12}k_{\nu}[\sigma(\partial_{\mu}\sigma)+\omega(\partial_{\mu}\omega)] \nonumber \\
&\qquad\qquad\left.+\,\frac{\rmi}{12}j^{\sigma}[\sigma(\partial_{\mu}\sdual_{\nu\sigma})+\omega(\partial_{\mu}s_{\nu\sigma})]-\frac{5\rmi}{12}(\partial_{\mu}j^{\sigma})(\sigma\sdual_{\nu\sigma}+\omega s_{\nu\sigma})\right\}.
\end{align}
This is entirely in terms of bilinears, but we would like to go further and eliminate the rank-2 terms, $s_{\mu\nu}$ and $\sdual_{\mu\nu}$. Using the known Fierz identities for the replacement of these terms \cite{Kaempffer1981}
\begin{align}
&s^{\mu\nu}=(\sigma^{2}-\omega^{2})^{-1}(\sigma\epsilon^{\mu\nu\rho\sigma}-\omega\delta^{\mu\nu\rho\sigma})j_{\rho}k_{\sigma}, \\
&\sdual^{\mu\nu}=(\sigma^{2}-\omega^{2})^{-1}(\omega\epsilon^{\mu\nu\rho\sigma}-\sigma\delta^{\mu\nu\rho\sigma})j_{\rho}k_{\sigma},
\end{align}
where we define the partially antisymmetric object
\begin{equation}
\delta^{\mu\nu\rho\sigma}\equiv\rmi(\eta^{\mu\rho}\eta^{\nu\sigma}-\eta^{\mu\sigma}\eta^{\nu\rho}),
\end{equation}
the rank-2 dependent terms in (\ref{Belinfante Fierz identity, with rank-2 terms}) become
\begin{align}
&\frac{\rmi}{12}j^{\sigma}[\sigma(\partial_{\mu}\sdual_{\nu\sigma})+\omega(\partial_{\mu}s_{\nu\sigma})]=\frac{1}{12}(\sigma^{2}-\omega^{2})^{-1}\left\{2\sigma\omega k_{\nu}[\omega(\partial_{\mu}\sigma)-\sigma(\partial_{\mu}\omega)]\right. \nonumber \\
&\left.{}\qquad+2\rmi\sigma\omega\epsilon_{\nu\sigma\rho\epsilon}j^{\sigma}(\partial_{\mu}j^{\rho})k^{\epsilon}+j_{\nu}j^{\sigma}(\partial_{\mu}k_{\sigma})(\sigma^{2}+\omega^{2})\right\} -\frac{1}{12}(\partial_{\mu}k_{\nu})(\sigma^{2}+\omega^{2}),
\end{align}
and
\begin{align}
&-\frac{5\rmi}{12}(\partial_{\mu}j^{\sigma})(\sigma\sdual_{\nu\sigma}+\omega s_{\nu\sigma})=(\sigma^{2}-\omega^{2})^{-1}\left\{\frac{5}{12}j_{\nu}j^{\sigma}(\partial_{\mu}k_{\sigma})(\sigma^{2}+\omega^{2})\right. \nonumber \\
&\left.{}\qquad-\frac{5\rmi}{6}\sigma\omega\epsilon_{\nu\sigma\rho\epsilon}(\partial_{\mu}j^{\sigma})j^{\rho}k^{\epsilon}-\frac{5}{12}k_{\nu}(\sigma^{2}+\omega^{2})[\omega(\partial_{\mu}\omega)-\sigma(\partial_{\mu}\sigma)]\right\},
\end{align}
giving us the final form of our identity
\begin{align}\label{Belinfante Fierz identity}
&[\psibar\gamma_{\nu}(\partial_{\mu}\psi)-(\partial_{\mu}\psibar)\gamma_{\nu}\psi]=(\sigma^{2}-\omega^{2})^{-1}\{k_{\nu}[\omega(\partial_{\mu}\sigma)-\sigma(\partial_{\mu}\omega)]-\rmi\epsilon_{\nu\sigma\rho\epsilon}(\partial_{\mu}j^{\sigma})j^{\rho}k^{\epsilon} \nonumber \\
&\qquad-\rmi j_{\nu}m^{\sigma}(\partial_{\mu}n_{\sigma})\}.
\end{align}
Substituting into (\ref{Belinfante stress-energy for free Dirac particle}) and relabelling some indices for convenience later on, we obtain the bilinear form of the Belinfante stress-energy tensor for a free Dirac particle
\begin{align}\label{Belinfante stress-energy tensor, free Dirac}
\Theta_{\mu\nu,\rmD}={}&\frac{1}{4}(\sigma^{2}-\omega^{2})^{-1}\{\rmi[k_{\mu}(\omega\partial_{\nu}\sigma-\sigma\partial_{\nu}\omega)+k_{\nu}(\omega\partial_{\mu}\sigma-\sigma\partial_{\mu}\omega)] \nonumber \\
&+j_{\sigma}k_{\kappa}[\epsilon_{\nu}{}^{\rho\sigma\kappa}(\partial_{\mu}j_{\rho})+\epsilon_{\mu}{}^{\rho\sigma\kappa}(\partial_{\nu}j_{\rho})]+j_{\mu}m^{\sigma}(\partial_{\nu}n_{\sigma})+j_{\nu}m^{\sigma}(\partial_{\mu}n_{\sigma})\}.
\end{align}
The Fierz identity (\ref{Belinfante Fierz identity}) is of central importance to our current work on the stress-energy tensor, since it permits an explicit mapping from the Dirac spinor fields to the physically equivalent set of bilinears, in the form in which the spinors appear in the Belinfante tensor (\ref{Belinfante stress-energy for free Dirac particle}). Of note is that none of the elements of the bilinear set (\ref{Real Dirac bilinears}) which we have chosen to work with contain any \emph{internal} derivative operators, so all gradient terms appear \emph{explicitly} in (\ref{Belinfante stress-energy tensor, free Dirac}). Additionally, (\ref{Belinfante Fierz identity}), along with (\ref{Anti-product rule on sigma}) and (\ref{Anti-product rule on omega}), are examples of a large family of related Fierz identities, and only represent the tip of the iceberg of such relations \cite{Takahashi1983}.

\subsection{Maxwell-Dirac Belinfante tensor}
The full Maxwell-Dirac stress-energy tensor is
\begin{equation}
\Theta_{\mu\nu,\rmMD}=\Theta_{\mu\nu,\rmD}+\Theta_{\mu\nu,\rmint}+\Theta_{\mu\nu,\rmem},
\end{equation}
where $\Theta_{\mu\nu,\rmint}$ and $\Theta_{\mu\nu,\rmem}$ are the interaction and Maxwell field contributions respectively. The Maxwell contribution has the well known form
\begin{equation}\label{Maxwell stress-energy tensor}
\Theta_{\mu\nu,\rmem}=\frac{1}{4}\eta_{\mu\nu}F_{\sigma\rho}F^{\sigma\rho}-F_{\mu\sigma}F_{\nu}{}^{\sigma},
\end{equation}
consistent with our metric signature $(+---)$, and the interaction term is
\begin{equation}\label{Interaction stress-energy}
\Theta_{\mu\nu,\rmint}=-\frac{q}{2}(j_{\mu}A_{\nu}+j_{\nu}A_{\mu}),
\end{equation}
where the electromagnetic vector potential $A_{\mu}$ can be replaced by the gauge independent analogue $B_{\mu}$ using the definition from \cite{InglisJarvis2014}
\begin{equation}\label{Gauge invariant vector potential definition}
A_{\mu}=B_{\mu}+\frac{1}{2q}(\sigma^{2}-\omega^{2})^{-1}m^{\sigma}(\partial_{\mu}n_{\sigma}).
\end{equation}
The gauge dependent bilinear terms in (\ref{Interaction stress-energy}) cancel out the corresponding terms in (\ref{Belinfante stress-energy tensor, free Dirac}) exactly, so the full Maxwell-Dirac Belinfante stress-energy tensor is
\begin{align}\label{Maxwell-Dirac Belinfante stress-energy tensor}
\Theta_{\mu\nu,\rmMD}={}&\frac{1}{4}(\sigma^{2}-\omega^{2})^{-1}\{\rmi[k_{\mu}(\omega\partial_{\nu}\sigma-\sigma\partial_{\nu}\omega)+k_{\nu}(\omega\partial_{\mu}\sigma-\sigma\partial_{\mu}\omega)] \nonumber \\
&+j_{\sigma}k_{\kappa}[\epsilon_{\nu}{}^{\rho\sigma\kappa}(\partial_{\mu}j_{\rho})\!+\!\epsilon_{\mu}{}^{\rho\sigma\kappa}(\partial_{\nu}j_{\rho})]\}-\frac{q}{2}(j_{\mu}B_{\nu}+j_{\nu}B_{\mu}) \nonumber \\
&+\frac{1}{4}\eta_{\mu\nu}F_{\sigma\rho}F^{\sigma\rho}-F_{\mu\sigma}F_{\nu}{}^{\sigma},
\end{align}
which is manifestly symmetric and gauge independent.

\section{Maxwell-Dirac stress-energy tensor via general relativity}\label{Maxwell-Dirac stress-energy tensor via general relativity}
\subsection{Bilinear form of Dirac Lagrangian}
We will now derive the bilinear form of the Maxwell-Dirac stress-energy tensor again, this time using a completely different method. The approach outlined in section \ref{Maxwell-Dirac stress-energy tensor via Belinfante} involved the use of Fierz identities to convert the spinorial form of the Belinfante stress-energy for a free Dirac particle (\ref{Belinfante stress-energy for free Dirac particle}) into bilinear form (\ref{Belinfante stress-energy tensor, free Dirac}), to which the known tensor forms of the interaction and Maxwell contributions were added, yielding (\ref{Maxwell-Dirac Belinfante stress-energy tensor}).

This time around, we convert the Lagrangian for an interacting Dirac particle
\begin{equation}\label{Electromagnetically interacting Dirac Lagrangian}
\sL=\frac{\rmi}{2}\left[\psibar\gamma^{\mu}(\partial_{\mu}\psi)-(\partial_{\mu}\psibar)\gamma^{\mu}\psi\right]-m\psibar\psi-q\psibar\gamma^{\mu}\psi A_{\mu},
\end{equation}
into bilinear form, then use the definition of the stress-energy tensor from general relativity to obtain our result, which should in principle agree with (\ref{Maxwell-Dirac Belinfante stress-energy tensor}). Note that this method from general relativity was used directly on the spinorial form of the Lagrangian (\ref{Free Dirac Lagrangian}) by Goedecke \cite{Goedecke1974}, who then demonstrated its equivalence with the Belinfante stress-energy tensor (\ref{Belinfante stress-energy for free Dirac particle}). We are pursuing a similar equivalence demonstration, with our focus being on the bilinear formalism. However, in contrast with the calculation of the previous section where the Fierz bilinear transcription was performed at the end, once the spinorial Belinfante tensor (\ref{Belinfante stress-energy for free Dirac particle}) had been obtained, here we transcribe \emph{at the start}, and perform the entire variational calculation with the bilinear field variables. The Fierz identity required to rewrite (\ref{Electromagnetically interacting Dirac Lagrangian}) is obtained by simply substituting the contracted form of (\ref{Belinfante Fierz identity}),
\begin{align}\label{Contracted Belinfante Fierz identity}
&[\psibar\gamma^{\mu}(\partial_{\mu}\psi)-(\partial_{\mu}\psibar)\gamma^{\mu}\psi]=(\sigma^{2}-\omega^{2})^{-1}\{k^{\mu}[\omega(\partial_{\mu}\sigma)-\sigma(\partial_{\mu}\omega)]-\rmi\epsilon^{\mu\sigma\rho\epsilon}(\partial_{\mu}j_{\sigma})j_{\rho}k_{\epsilon} \nonumber \\
&\qquad-\rmi j^{\mu}m^{\sigma}(\partial_{\mu}n_{\sigma})\}.
\end{align}
Applying the definitions $\sigma\equiv\psibar\psi$, $j^{\mu}\equiv\psibar\gamma^{\mu}\psi$, and using the definition of $B_{\mu}$ (\ref{Gauge invariant vector potential definition}), we obtain
\begin{align}\label{Interacting Dirac Lagrangian in bilinear form}
\sL=\frac{1}{2}(\sigma^{2}-\omega^{2})^{-1}\{\rmi k^{\rho}[\omega(\partial_{\rho}\sigma)-\sigma(\partial_{\rho}\omega)]+\epsilon^{\rho\sigma\kappa\tau}(\partial_{\rho}j_{\sigma})j_{\kappa}k_{\tau}\}-m\sigma-qj^{\rho}B_{\rho}.
\end{align}

\subsection{Variational form of the stress-energy tensor}
The total action for the gravitational field in the presence of matter is \cite{Wald1984}
\begin{equation}\label{Action in presence of matter}
S=\frac{S_{\rmH}}{16\pi G}+S_{\rmM},
\end{equation}
where $S_{\rmM}$ is the action for matter fields (mass-energy). $S_{\rmH}$ is the Hilbert action, defined as
\begin{equation}\label{Hilbert action}
S_{\rmH}=\int\rmd^{4}x\sqrt{-g}R,
\end{equation}
where $g$ is the determinant of the metric $g_{\mu\nu}$, $R$ is the Ricci scalar, and $\rmd^{4}x$ is the invariant volume element. The variation of the action with respect to an arbitrary tensor field $\Phi^{\mu_{1}...\mu_{k}}{}_{\nu_{1}...\nu_{l}}$ takes the general form
\begin{equation}\label{General action variational form}
\delta S=\int\rmd^{4}x\frac{\delta S}{\delta\Phi}\delta\Phi,
\end{equation}
with contraction over the indices implied. The term $\delta S/\delta\Phi$ is called the functional derivative of $S$ with respect to the tensor field $\Phi$. Of main interest in variational theory are tensors $\Phi_{0}$ which extremize the action, so that $\delta S=0$, and hence
\begin{equation}
\frac{\delta S}{\delta\Phi}\bigg|_{\Phi_{0}}=0.
\end{equation}
Extremizing the variation of the Hilbert action (\ref{Hilbert action}) with respect to the inverse metric leads to Einstein's equations in vacuum
\begin{equation}
\frac{1}{\sqrt{-g}}\frac{\delta S_{\rmH}}{\delta g^{\mu\nu}}=R_{\mu\nu}-\frac{1}{2}R g_{\mu\nu}=0.
\end{equation}
Likewise, extremizing the gravitational action in the presence of matter (\ref{Action in presence of matter}), so that
\begin{equation}
\delta S=\frac{\delta S_{\rmH}}{16\pi G}+\delta S_{\rmM}=0,
\end{equation}
and equating the corresponding functional derivatives, yields
\begin{equation}
\frac{1}{\sqrt{-g}}\frac{\delta S}{\delta g^{\mu\nu}}=\frac{1}{16\pi G}\left(R_{\mu\nu}-\frac{1}{2}Rg_{\mu\nu}\right)+\frac{1}{\sqrt{-g}}\frac{\delta S_{\rmM}}{\delta g^{\mu\nu}}=0.
\end{equation}
Comparing with the well-known form of Einstein's equations in the presence of matter
\begin{equation}
R_{\mu\nu}-\frac{1}{2}Rg_{\mu\nu}=8\pi GT_{\mu\nu},
\end{equation}
we can see that the energy momentum tensor is of the general form
\begin{equation}\label{Variational stress-energy tensor}
T_{\mu\nu}=-\frac{2}{\sqrt{-g}}\frac{\delta S_{\rmM}}{\delta g^{\mu\nu}}.
\end{equation}

\subsection{Variational Maxwell-Dirac stress-energy tensor}
The variation of the electromagnetically interacting Dirac matter action, ignoring the contribution of the Maxwell field itself for now, is given by
\begin{equation}\label{Variation of action, initial form}
\delta S=\int\rmd^{4}x\left(\delta\sqrt{-g}\sL+\sqrt{-g}\delta\sL\right),
\end{equation}
where the Lagrangian is given by (\ref{Interacting Dirac Lagrangian in bilinear form}), but with appropriate modifications to make it covariant in curved space. Since the volume element $\rmd^{4}x$ and $\sqrt{-g}$ are scalar densities of weight $-1$ and $+1$ respectively, we must arrange for the Lagrangian to be manifestly a scalar. Notice that (\ref{Interacting Dirac Lagrangian in bilinear form}) contains a term dependent on the Levi-Civita symbol with upstairs indices, which is of weight $-1$. This implies that we should make the replacement
\begin{equation}
\epsilon^{\rho\sigma\kappa\tau}\rightarrow\frac{1}{\sqrt{-g}}\epsilon^{\rho\sigma\kappa\tau}.
\end{equation}
In order to deal with the bilinear four-vectors we must introduce the vierbein fields \cite{Yepez2011}, which locally relate the curved metric to the flat one
\begin{equation}\label{Metric in terms of vierbeins}
g_{\mu\nu}=e_{\mu}{}^{a}e_{\nu}{}^{b}\eta_{ab},
\end{equation}
where Greek and Latin indices label curved and flat spacetime components respectively. The gamma matrices are modified such that
\begin{equation}
\gamma_{\mu}=e_{\mu}{}^{a}\gamma_{a};\qquad\{\gamma_{a},\gamma_{b}\}=2\eta_{ab},
\end{equation}
so the bilinears are now
\begin{align}
&j_{\mu}=e_{\mu}{}^{a}j_{a};\qquad j_{a}=\psibar\gamma_{a}\psi, \\
&k_{\mu}=e_{\mu}{}^{a}k_{a};\qquad k_{a}=\psibar\gamma_{5}\gamma_{a}\psi.
\end{align}
The variation of the square root of the negative metric determinant is \cite{Wald1984}
\begin{equation}
\delta\sqrt{-g}=-\frac{1}{2}\sqrt{-g}\,g_{\mu\nu}\delta g^{\mu\nu}.
\end{equation}
Noting that $h=\sqrt{-g}$, where $h$ is the vierbein determinant, we can use the variation of (\ref{Metric in terms of vierbeins}) to alternatively write this as
\begin{equation}
\delta h=-he_{\mu}{}^{a}(\delta e^{\mu}{}_{a}),
\end{equation}
implying the reciprocal variation
\begin{equation}
\delta(h^{-1})=h^{-1}e_{\mu}{}^{a}(\delta e^{\mu}{}_{a}).
\end{equation}
In curved space, the Levi-Civita term in (\ref{Interacting Dirac Lagrangian in bilinear form}) becomes
\begin{equation}
h^{-1}\epsilon^{\rho\sigma\kappa\tau}(\partial_{\rho}e_{\sigma}{}^{a})j_{a}j_{\kappa}k_{\tau}+h^{-1}\epsilon^{\rho\sigma\kappa\tau}e_{\sigma}{}^{a}(\partial_{\rho}j_{a})j_{\kappa}k_{\tau}.
\end{equation}
Introducing the covariant derivative causes the first term to vanish, due to the tetrad postulate \cite{Yepez2011}
\begin{equation}
\nabla_{\mu}e_{\nu}{}^{a}=0.
\end{equation}
Expanding out all of the vierbein fields in the second term, we find that
\begin{equation}
\epsilon^{\rho\sigma\kappa\tau}e_{\sigma}{}^{a}(\partial_{\rho}j_{a})j_{\kappa}k_{\tau}=\epsilon^{abcd}(\partial_{a}j_{b})j_{c}k_{d},
\end{equation}
which implies that for any curved coordinate components, this term is always equal to the flat spacetime version, so it is automatically covariant. We find that the covariant bilinear electromagnetically interacting Dirac matter Lagrangian has the form
\begin{align}
\sL={}&\frac{1}{2}(\sigma^{2}-\omega^{2})^{-1}[\rmi g^{\rho\sigma}e_{\sigma}{}^{a}k_{a}(\omega\partial_{\rho}\sigma-\sigma\partial_{\rho}\omega)+h^{-1}\epsilon^{\rho\sigma\kappa\tau}e_{\sigma}{}^{a}e_{\kappa}{}^{b}e_{\tau}{}^{c}(\partial_{\rho}j_{a})j_{b}k_{c}] \nonumber \\
&-m\sigma-qg^{\rho\sigma}e_{\sigma}{}^{a}j_{a}B_{\rho}.
\end{align}
The variation with respect to deformation of the vierbein field is
\begin{align}\label{Lagrangian variation, before simplification}
\delta\sL={}&\frac{1}{2}(\sigma^{2}-\omega^{2})^{-1}\{\rmi\delta(g^{\rho\sigma}e_{\sigma}{}^{a})k_{a}(\omega\partial_{\rho}\sigma-\sigma\partial_{\rho}\omega)+\epsilon^{\rho\sigma\kappa\tau}\delta(h^{-1}e_{\sigma}{}^{a}e_{\kappa}{}^{b}e_{\tau}{}^{c})(\partial_{\rho}j_{a})j_{b}k_{c}\} \nonumber \\
&-q\delta(g^{\rho\sigma}e_{\sigma}{}^{a})j_{a}B_{\rho}.
\end{align}
From the variation of (\ref{Metric in terms of vierbeins}), we find that
\begin{equation}
\delta(g^{\rho\sigma}e_{\sigma}{}^{a})=2(\delta e^{\rho a})+(\delta e^{\sigma}{}_{b})e^{\rho b}e_{\sigma}{}^{a}.
\end{equation}
Using the fundamental vierbein property
\begin{align}
&e_{\mu}{}^{a}e^{\mu}{}_{b}=\delta^{a}_{b}, \\
&e_{\mu}{}^{a}(\delta e^{\mu}{}_{b})=-(\delta e_{\mu}{}^{a})e^{\mu}{}_{b},
\end{align}
we find that the first and last terms in (\ref{Lagrangian variation, before simplification}) are
\begin{align}
&\rmi\delta(g^{\rho\sigma}e_{\sigma}{}^{a})k_{a}(\omega\partial_{\rho}\sigma-\sigma\partial_{\rho}\omega)=\rmi k^{a}(\omega\partial_{\mu}\sigma-\sigma\partial_{\mu}\omega)(\delta e^{\mu}{}_{a}), \\
&-q\delta(g^{\rho\sigma}e_{\sigma}{}^{a})j_{a}B_{\rho}=-qj^{a}B_{\mu}(\delta e^{\mu}{}_{a}).
\end{align}
Following a similar process, we find that the second variational term is
\begin{align}
&\epsilon^{\rho\sigma\kappa\tau}\delta(h^{-1}e_{\sigma}{}^{a}e_{\kappa}{}^{b}e_{\tau}{}^{c})(\partial_{\rho}j_{a})j_{b}k_{c} \nonumber \\
&\qquad=h^{-1}[e_{\mu}{}^{a}\epsilon^{\rho\sigma\kappa\tau}e_{\sigma}{}^{d}(\partial_{\rho}j_{d})j_{\kappa}k_{\tau}-\epsilon^{\rho}{}_{\nu}{}^{\sigma\kappa}(\partial_{\rho}j_{b})j_{\sigma}k_{\kappa}e^{\nu a}e_{\mu}{}^{b}-\epsilon^{\rho\sigma}{}_{\nu}{}^{\kappa}(\partial_{\rho}j_{b})j_{\mu}k_{\kappa}e^{\nu a}e_{\sigma}{}^{b} \nonumber \\
&\qquad\qquad-\epsilon^{\rho\sigma\kappa}{}_{\nu}(\partial_{\rho}j_{b})j_{\kappa}k_{\mu}e^{\nu a}e_{\sigma}{}^{b}](\delta e^{\mu}{}_{a}).
\end{align}
Gathering the deformed terms together, we can write the variation of the Lagrangian as
\begin{align}
\delta\sL={}&\left(\frac{1}{2}(\sigma^{2}-\omega^{2})^{-1}\{\rmi k^{a}(\omega\partial_{\mu}\sigma-\sigma\partial_{\mu}\omega)+h^{-1}[e_{\mu}{}^{a}\epsilon^{\rho\sigma\kappa\tau}e_{\sigma}{}^{d}(\partial_{\rho}j_{d})j_{\kappa}k_{\tau}\right. \nonumber \\
&-\epsilon^{\rho}{}_{\nu}{}^{\sigma\kappa}(\partial_{\rho}j_{b})j_{\sigma}k_{\kappa}e^{\nu a}e_{\mu}{}^{b}-\epsilon^{\rho\sigma}{}_{\nu}{}^{\kappa}(\partial_{\rho}j_{b})j_{\mu}k_{\kappa}e^{\nu a}e_{\sigma}{}^{b}-\epsilon^{\rho\sigma\kappa}{}_{\nu}(\partial_{\rho}j_{b})j_{\kappa}k_{\mu}e^{\nu a}e_{\sigma}{}^{b}]\} \nonumber \\
&\left.-qj^{a}B_{\mu}\right)(\delta e^{\mu}{}_{a})
\end{align}
with the associated action variation being
\begin{align}
\delta S_{\rmD}={}&\int\rmd^{4}x\sqrt{-g}\left(-e_{\mu}{}^{a}\left\{\frac{1}{2}(\sigma^{2}-\omega^{2})^{-1}[\rmi k^{\rho}(\omega\partial_{\rho}\sigma-\sigma\partial_{\rho}\omega)+\epsilon^{\rho\sigma\kappa\tau}(\partial_{\rho}j_{\sigma})j_{\kappa}k_{\tau}]\right.\right. \nonumber \\
&\left.-m\sigma-qj^{\rho}B_{\rho}\right\}+\frac{1}{2}(\sigma^{2}-\omega^{2})^{-1}\{\rmi k^{a}(\omega\partial_{\mu}\sigma-\sigma\partial_{\mu}\omega) \nonumber \\
&+h^{-1}[e_{\mu}{}^{a}\epsilon^{\rho\sigma\kappa\tau}e_{\sigma}{}^{d}(\partial_{\rho}j_{d})j_{\kappa}k_{\tau}-\epsilon^{\rho}{}_{\nu}{}^{\sigma\kappa}(\partial_{\rho}j_{b})j_{\sigma}k_{\kappa}e^{\nu a}e_{\mu}{}^{b} \nonumber \\
&\left.{}-\epsilon^{\rho\sigma}{}_{\nu}{}^{\kappa}(\partial_{\rho}j_{b})j_{\mu}k_{\kappa}e^{\nu a}e_{\sigma}{}^{b}-\epsilon^{\rho\sigma\kappa}{}_{\nu}(\partial_{\rho}j_{b})j_{\kappa}k_{\mu}e^{\nu a}e_{\sigma}{}^{b}]\}-qj^{a}B_{\mu}\right)(\delta e^{\mu}{}_{a}).\label{Action variation with Lagrangian variation substituted}
\end{align}
From the general form of the action variation (\ref{General action variational form}), a relationship between (\ref{Action variation with Lagrangian variation substituted}) and the stress-energy tensor can be obtained \cite{Yepez2011}
\begin{equation}
\delta S_{\rmD}=\int\rmd^{4}x\sqrt{-g}\,u_{\lambda}{}^{a}\delta e^{\lambda}{}_{a}=\frac{1}{2}\int\rmd^{4}x\sqrt{-g}\,T^{\mu\nu}\delta g_{\mu\nu},
\end{equation}
which implies that
\begin{equation}
u_{\mu}{}^{a}=\frac{1}{2}(T_{\mu\lambda}e^{\lambda a}+T_{\lambda\mu}e^{\lambda a}).
\end{equation}
Recognizing $T_{\mu\nu}$ as symmetric gives
\begin{equation}
T_{\mu\nu}=\frac{1}{2}(e_{\mu a}u_{\nu}{}^{a}+e_{\nu a}u_{\mu}{}^{a}).
\end{equation}
Identifying the contents of the external parentheses in (\ref{Action variation with Lagrangian variation substituted}) with $u_{\mu}{}^{a}$, we obtain for the stress energy tensor
\begin{align}
&T_{\mu\nu}=-\eta_{\mu\nu}\left\{\frac{1}{2}(\sigma^{2}-\omega^{2})^{-1}[\rmi k^{\rho}(\omega\partial_{\rho}\sigma-\sigma\partial_{\rho}\omega)+\epsilon^{\rho\sigma\kappa\tau}(\partial_{\rho}j_{\sigma})j_{\kappa}k_{\tau}]-m\sigma-qj^{\rho}B_{\rho}\right\} \nonumber \\
&\ +\frac{1}{4}(\sigma^{2}-\omega^{2})^{-1}\{\rmi[k_{\mu}(\omega\partial_{\nu}\sigma-\sigma\partial_{\nu}\omega)+k_{\nu}(\omega\partial_{\mu}\sigma-\sigma\partial_{\mu}\omega)]+2\eta_{\mu\nu}\epsilon^{\rho\sigma\kappa\tau}(\partial_{\rho}j_{\sigma})j_{\kappa}k_{\tau} \nonumber \\
&\ -\epsilon^{\rho}{}_{\mu}{}^{\sigma\kappa}(\partial_{\rho}j_{\nu})j_{\sigma}k_{\kappa}-\epsilon^{\rho\sigma}{}_{\mu}{}^{\kappa}(\partial_{\rho}j_{\sigma})j_{\nu}k_{\kappa}-\epsilon^{\rho\sigma\kappa}{}_{\mu}(\partial_{\rho}j_{\sigma})j_{\kappa}k_{\nu}-\epsilon^{\rho}{}_{\nu}{}^{\sigma\kappa}(\partial_{\rho}j_{\mu})j_{\sigma}k_{\kappa} \nonumber \\
&\ -\epsilon^{\rho\sigma}{}_{\nu}{}^{\kappa}(\partial_{\rho}j_{\sigma})j_{\mu}k_{\kappa}-\epsilon^{\rho\sigma\kappa}{}_{\nu}(\partial_{\rho}j_{\sigma})j_{\kappa}k_{\mu}\}-\frac{q}{2}(j_{\mu}B_{\nu}+j_{\nu}B_{\mu}),\label{Stress-energy, with flat metric term}
\end{align}
where we have evaluated at flat spacetime. This is manifestly symmetric, but it requires some additional manipulation before it more closely resembles the Belinfante form (\ref{Maxwell-Dirac Belinfante stress-energy tensor}). Consider the $U(1)$ gauge covariant Dirac equation and its Dirac conjugate
\begin{align}
&\rmi\gamma^{\sigma}(\partial_{\sigma}\psi)-q\gamma^{\sigma}A_{\sigma}\psi-m\psi=0, \\
&\rmi(\partial_{\sigma}\psibar)\gamma^{\sigma}+q\psibar\gamma^{\sigma}A_{\sigma}+m\psibar=0.
\end{align}
Left and right multiplying these equations by $\psibar$ and $\psi$ respectively, then subtracting the second from the first and rearranging, gives
\begin{equation}
m\sigma=\frac{\rmi}{2}\left[\psibar\gamma^{\sigma}(\partial_{\sigma}\psi)-(\partial_{\sigma}\psibar)\gamma^{\sigma}\psi\right]-qj^{\sigma}A_{\sigma}
\end{equation}
Applying the Fierz identity (\ref{Contracted Belinfante Fierz identity}) and the $B_{\mu}$ definition (\ref{Gauge invariant vector potential definition}), this becomes
\begin{align}
-m\sigma={}&-\frac{1}{2}(\sigma^{2}-\omega^{2})^{-1}[\rmi k^{\rho}(\omega\partial_{\rho}\sigma-\sigma\partial_{\rho}\omega)+\epsilon^{\rho\sigma\kappa\tau}(\partial_{\rho}j_{\sigma})j_{\kappa}k_{\tau}]+qj^{\rho}B_{\rho},
\end{align}
causing the $\eta_{\mu\nu}$ dependent term in (\ref{Stress-energy, with flat metric term}) to vanish. Now consider the combinatorial identity\footnote{This follows from the 5 term cyclic identity $V^{\alpha}\epsilon^{\rho\sigma\kappa\tau}+V^{\tau}\epsilon^{\alpha\rho\sigma\kappa}+\cdots=0$ which holds for the Levi-Civita tensor multiplied by any contravariant vector quantity. With the role of $V^{\alpha}$ played by the Kronecker $\delta^{\alpha}_{\beta}$ (for fixed $\beta$), this yields (\ref{Levi-Civita combinatorial identity}) after contracting with $\eta_{\mu\alpha}\delta^{\beta}_{\nu}(\partial_{\rho}j_{\sigma})j_{\kappa}k_{\tau}$, and rearranging indices appropriately.}
\begin{align}
&\epsilon_{\mu}{}^{\rho\sigma\kappa}(\partial_{\nu}j_{\rho})j_{\sigma}k_{\kappa}+\epsilon_{\nu}{}^{\rho\sigma\kappa}(\partial_{\mu}j_{\rho})j_{\sigma}k_{\kappa} \nonumber \\
&\qquad=2\eta_{\mu\nu}\epsilon^{\rho\sigma\kappa\tau}(\partial_{\rho}j_{\sigma})j_{\kappa}k_{\tau}-\epsilon^{\rho}{}_{\mu}{}^{\sigma\kappa}(\partial_{\rho}j_{\nu})j_{\sigma}k_{\kappa}-\epsilon^{\rho\sigma}{}_{\mu}{}^{\kappa}(\partial_{\rho}j_{\sigma})j_{\nu}k_{\kappa}-\epsilon^{\rho\sigma\kappa}{}_{\mu}(\partial_{\rho}j_{\sigma})j_{\kappa}k_{\nu} \nonumber \\
&\qquad\qquad-\epsilon^{\rho}{}_{\nu}{}^{\sigma\kappa}(\partial_{\rho}j_{\mu})j_{\sigma}k_{\kappa}-\epsilon^{\rho\sigma}{}_{\nu}{}^{\kappa}(\partial_{\rho}j_{\sigma})j_{\mu}k_{\kappa}-\epsilon^{\rho\sigma\kappa}{}_{\nu}(\partial_{\rho}j_{\sigma})j_{\kappa}k_{\mu},\label{Levi-Civita combinatorial identity}
\end{align}
which can be used to obtain the final form of the variational stress-energy tensor for Dirac matter
\begin{align}
T_{\mu\nu,\rmD}={}&\frac{1}{4}(\sigma^{2}-\omega^{2})^{-1}\{\rmi[k_{\mu}(\omega\partial_{\nu}\sigma-\sigma\partial_{\nu}\omega)+k_{\nu}(\omega\partial_{\mu}\sigma-\sigma\partial_{\mu}\omega)] \nonumber \\
&+j_{\sigma}k_{\kappa}[\epsilon_{\nu}{}^{\rho\sigma\kappa}(\partial_{\mu}j_{\rho})+\epsilon_{\mu}{}^{\rho\sigma\kappa}(\partial_{\nu}j_{\rho})]\}-\frac{q}{2}(j_{\mu}B_{\nu}+j_{\nu}B_{\mu}).
\end{align}
Comparing with the Belinfante tensor (\ref{Maxwell-Dirac Belinfante stress-energy tensor}), we find that they agree
\begin{equation}
T_{\mu\nu,\rmMD}=\Theta_{\mu\nu,\rmMD},
\end{equation}
when the gauge field stress-energy (\ref{Maxwell stress-energy tensor}) is included on the left-hand side.

\section{Symmetry reduction of the Maxwell-Dirac stress-energy tensor}\label{Symmetry reduction of the Maxwell-Dirac stress-energy tensor}
The Maxwell-Dirac equations in the bilinear formalism are in general, a very complicated set of self-coupled partial differential equations. An application of the present construction of the physical stress-energy tensor of the system in terms of bilinears, is therefore to provide a representation of the conserved rest mass of possible solutions (via the spatial integral of $T_{00}$ for example).

For any meaningful solutions to be derived, it is natural to consider reduced forms under the imposition of special symmetries. The reduction of the bilinear form of the Maxwell-Dirac system under several examples of subgroups of the Poincar\'{e} group was discussed in \cite{InglisJarvis2014}. We therefore choose one of the most important of these subgroups to work with here, namely $SO(3)$, which corresponds to spherical symmetry. In particular, we shall demonstrate how the bilinear form of the Maxwell-Dirac stress-energy tensor reduces, given the restrictions imposed by this subgroup. The treatment of other symmetry reductions, such as cylindrical symmetry and the $\tP_{13,10}$ subgroup from \cite{PWZ1975}, shall be left for future work.



Under spherical symmetry, scalar fields $(\sigma, \omega, \mathrm{etc.})$ have the generic form
\begin{equation}
\phi=f(t,r)
\end{equation}
and vector fields $(j^{\mu}, k^{\mu}, \mathrm{etc.})$ have the form
\begin{equation}
\Phi^{\mu}=\left(\begin{array}{c}f(t,r) \\
xg(t,r) \\
yg(t,r) \\
zg(t,r)\end{array}\right),
\end{equation}
where the invariant
\begin{equation}
r=\sqrt{x^{2}+y^{2}+z^{2}}
\end{equation}
is simply the spatial radius. We showed in \cite{InglisJarvis2014} that the spherically symmetric forms of our bilinear vector and axial vector fields are
\begin{equation}
j^{\mu}=\left(\begin{array}{c}j_{a} \\
xj_{b} \\
yj_{b} \\
zj_{b}\end{array}\right),\qquad k^{\mu}=\left(\begin{array}{c}rj_{b} \\
(x/r)j_{a} \\
(y/r)j_{a} \\
(z/r)j_{a}\end{array}\right),\qquad B^{\mu}=\left(\begin{array}{c}B_{a} \\
xB_{b} \\
yB_{b} \\
zB_{b}\end{array}\right),
\end{equation}
where the vector potential functions are
\begin{align}
&B_{a}=\left[\pm\frac{\rmi}{2}(\sigma_{r}\omega-\sigma\omega_{r})-m\sigma j_{a}\right]\left[q(\sigma^{2}-\omega^{2})\right]^{-1}, \\
&B_{b}=\left[\mp\frac{\rmi}{2r}(\sigma_{t}\omega-\sigma\omega_{t})-m\sigma j_{b}\right]\left[q(\sigma^{2}-\omega^{2})\right]^{-1}.
\end{align}
Here, we are using a condensed derivative notation $\partial_{t}\sigma\equiv\sigma_{t}$, and so on. Note that the effect of the symmetry reduction has in this case, reduced the components of the four-vectors to the set of coefficient functions $j_{a}$, $j_{b}$, $\sigma$ and $\omega$, which are themselves further constrained by higher-order nonlinear PDEs in the Maxwell-Dirac system. The coefficient functions $k_{a}$ and $k_{b}$ have been eliminated through the use of the Fierz identities
\begin{align}
&j_{\mu}j^{\mu}=-k_{\mu}k^{\mu}=\sigma^{2}-\omega^{2}, \\
&j_{\mu}k^{\mu}=0.
\end{align}
It is straightforward to show that the Levi-Civita terms in the stress-energy vanish in this symmetry case
\begin{equation}
j_{\sigma}k_{\kappa}[\epsilon_{\nu}{}^{\rho\sigma\kappa}(\partial_{\mu}j_{\rho})+\epsilon_{\mu}{}^{\rho\sigma\kappa}(\partial_{\nu}j_{\rho})]=0.
\end{equation}
The form of the stress-energy tensor we are dealing with is therefore
\begin{align}\label{Stress-energy tensor with spherical symmetry simplification}
T_{\mu\nu,\rmMD}={}&\frac{\rmi}{4}(\sigma^{2}-\omega^{2})^{-1}[k_{\mu}(\omega\partial_{\nu}\sigma-\sigma\partial_{\nu}\omega)+k_{\nu}(\omega\partial_{\mu}\sigma-\sigma\partial_{\mu}\omega)]-\frac{q}{2}(j_{\mu}B_{\nu}+j_{\nu}B_{\mu}) \nonumber \\
&+\frac{1}{4}\eta_{\mu\nu}F_{\sigma\rho}F^{\sigma\rho}-F_{\mu\sigma}F_{\nu}{}^{\sigma}.
\end{align}
For the $SO(3)$ symmetry, the components of the electromagnetic field strength tensor are
\begin{align}
&F_{0i}=x_{i}F_{a}(t,r), \\
&F_{ij}=\epsilon_{ijk}x^{k}F_{b}(t,r),
\end{align}
where the Maxwell coefficient functions are
\begin{align}
F_{a}(t,r)={}&(1/qr)(\sigma^{2}-\omega^{2})^{-2}\{-2m[\sigma j_{a}(\sigma\sigma_{r}-\omega\omega_{r})+r\sigma j_{b}(\sigma\sigma_{t}-\omega\omega_{t})] \nonumber \\
&\pm\rmi[\sigma\omega(\sigma_{r}^{2}-\sigma_{t}^{2}+\omega_{r}^{2}-\omega_{t}^{2})+(\sigma^{2}+\omega^{2})(\sigma_{t}\omega_{t}-\sigma_{r}\omega_{r})]\} \nonumber \\
&+(1/qr)(\sigma^{2}-\omega^{2})^{-1}[m(\sigma_{r}j_{a}+\sigma j_{a,r}+r\sigma_{t}j_{b}+r\sigma j_{b,t}) \nonumber \\
&\pm(\rmi/2)(\sigma_{tt}\omega-\sigma\omega_{tt}-\sigma_{rr}\omega+\sigma\omega_{rr})],
\end{align}
representing the electric field form, and
\begin{equation}
F_{b}(r)=\pm\frac{1}{2qr^{3}},
\end{equation}
representing the magnetic field form, which happens to be that of a monopole. Treating the $\mu=\nu=0$, $\mu=0$, $\nu=i$ and $\mu=i$, $\nu=j$ cases separately, we find that the respective components of (\ref{Stress-energy tensor with spherical symmetry simplification}) are
\begin{align}
&T_{00,\rmMD}=T_{a}+\scF, \\
&T_{0i,\rmMD}=\frac{x_{i}}{r}T_{b}, \\
&T_{ij,\rmMD}=\frac{x_{i}x_{j}}{r^{2}}T_{c}+\delta_{ij}\scF,
\end{align}
where the energy density of the Maxwell field is
\begin{equation}\label{Spherical Maxwell stress-energy}
\scF=\frac{r^{2}(F_{a}^{2}+F_{b}^{2})}{2},
\end{equation}
and the other functions are defined as
\begin{align}
&T_{a}(t,r)=(\sigma^{2}-\omega^{2})^{-1}\left\{\pm\frac{\rmi}{2}\left[rj_{b}(\sigma_{t}\omega-\sigma\omega_{t})-j_{a}(\sigma_{r}\omega-\sigma\omega_{r})\right]+m\sigma j_{a}^{2}\right\},\label{Spherical stress-energy function a} \\
&T_{b}(t,r)=(\sigma^{2}-\omega^{2})^{-1}\left\{\pm\frac{\rmi}{2}\left[rj_{b}(\sigma_{r}\omega-\sigma\omega_{r})-j_{a}(\sigma_{t}\omega-\sigma\omega_{t})\right]-m\sigma j_{a}rj_{b}\right\},\label{Spherical stress-energy function b} \\
&T_{c}(t,r)=(\sigma^{2}-\omega^{2})^{-1}\left\{\pm\frac{\rmi}{2}\left[rj_{b}(\sigma_{t}\omega-\sigma\omega_{t})-j_{a}(\sigma_{r}\omega-\sigma\omega_{r})\right]+m\sigma r^{2}j_{b}^{2}\right\}-2\scF.\label{Spherical stress-energy function c}
\end{align}

\section{Conclusions}
The central aim of our current work was to obtain a form of the stress-energy tensor for the self-coupled Maxwell-Dirac system, in terms of bilinears. The motivation for working in the bilinear formalism in general, is given by the fact that spinor fields do not correspond directly to observables, due to their dependence on an arbitrary unobservable phase factor. On the other hand, the system of Dirac bilinears corresponds to ``density fields'', or ``probability distributions'', which are observable according to a fundamental postulate of quantum mechanics. However, the attractiveness of the new physical insights that could be gained by studying the fully non-linear quantum electrodynamics, in terms of the observable tensor fields given by bilinears \cite{Takabayasi1957}, is offset somewhat by the sheer complexity of the self-coupled system. Reducing this system under a symmetry subgroup of the Poincar\'{e} group can simplify it to a point where it is mathematically tractable, and solutions can potentially be obtained. Such solutions would have associated mass-energy and momentum fluxes, which can be calculated directly using a bilinear-dependent form of the symmetry reduced stress-energy tensor.

In sections \ref{Maxwell-Dirac stress-energy tensor via Belinfante} and \ref{Maxwell-Dirac stress-energy tensor via general relativity}, we demonstrated that bilinear forms for the stress-energy tensor ($\Theta_{\mu\nu}$ and $T_{\mu\nu}$ respectively) can indeed be calculated, by applying Fierz identities to the spinor terms appearing in the Belinfante and variational general relativistic calculational schemes. In the Belinfante case, the Fierz mapping was applied to the spinorial Belinfante tensor, and in the variational case, it was applied to the spinorial Lagrangian prior to the vierbein deformation. Despite the fact that these two methods are quite independent, with very different starting points from the point of view of the Fierz bilinear transcription, they are in agreement in accordance with Goedecke's conjecture \cite{Goedecke1974} and Lord's subsequent equivalence proof \cite{Lord1976}. This mathematical consistency adds weight to the validity of the bilinear representation of the Maxwell-Dirac system, and our corresponding bilinear stress-energy result.

However, there is a point of view from which this automatic agreement is somewhat surprising. When taking into consideration the details of the functional Jacobian corresponding to the spinor to bilinear mapping, one would expect there to be extra constraint terms entering into the bilinearized Lagrangian, with the lack of such terms in (\ref{Interacting Dirac Lagrangian in bilinear form}) leading to a disagreement with the Belinfante tensor in the \emph{bilinear} representation. A transcription of spinor electrodynamics into gauge invariant quantities in this spirit, has been given in the functional formalism by Rudolph and Kijowski \cite{KijowskiRudolph1993}, \cite{KijowskiRudolphRudolph1995}. In their bosonic transcription, Green's functions are given as functional integrals in whose integrands there are always additional accompanying field-dependent factors, and so an effective bosonic, local, purely Lagrangian formulation is not obtained. The details of the agreement between $\Theta_{\mu\nu}$ and $T_{\mu\nu}$ for the bilinear case, although highly encouraging, remains a matter deserving of further study.

Putting these technical concerns aside, we then turned to an example to demonstrate how the bilinear stress-energy tensor is reduced under spherical symmetry, using the generic $SO(3)$ invariant forms for scalar and four-vector fields discussed in \cite{InglisJarvis2014}. We found that the stress-energy components could be described in terms of three functions (\ref{Spherical stress-energy function a})-(\ref{Spherical stress-energy function c}) corresponding to the interacting Dirac matter contribution, as well as a single function (\ref{Spherical Maxwell stress-energy}), corresponding to the energy density of the Maxwell field.

The bilinear form of the Maxwell-Dirac stress-energy tensor (\ref{Maxwell-Dirac Belinfante stress-energy tensor}) may be easily applied to \emph{solutions} of the symmetry reduced Maxwell-Dirac equations, such as considered in \cite{InglisJarvis2014}, by performing the corresponding symmetry reduction on $T_{\mu\nu}$, as done in section \ref{Symmetry reduction of the Maxwell-Dirac stress-energy tensor} for the $SO(3)$ group, and directly substituting the solution fields. We intend to investigate both numerical and closed-form solutions explicitly, for the static spherically symmetric and the algebraic splitting groups $\tP_{13,10}$ \cite{InglisJarvis2014}, \cite{PWZ1975}, in follow-up works.

\section*{Acknowledgements}
This study was financially supported by the Australian Postgraduate Awards (APA) program.

\appendix
\section{Dirac identities}
\begin{align}
&\{\gamma^{\mu},\gamma^{\nu}\}=2\eta^{\mu\nu} \\
&[\gamma^{\mu},\gamma^{\nu}]=-2\rmi\sigma^{\mu\nu} \\
&\gamma^{5}=\gamma_{5}=-(\rmi/4!)\epsilon_{\mu\nu\rho\sigma}\gamma^{\mu}\gamma^{\nu}\gamma^{\rho}\gamma^{\sigma}=\rmi\gamma^{0}\gamma^{1}\gamma^{2}\gamma^{3}=-\rmi\gamma_{0}\gamma_{1}\gamma_{2}\gamma_{3} \\
&\gamma_{5}^{2}=I \\
&\{\gamma_{5},\gamma^{\mu}\}=0 \\
&[\gamma_{5},\sigma^{\mu\nu}]=0 \\
&\gamma^{\mu}\gamma^{\nu}=\eta^{\mu\nu}-\rmi\sigma^{\mu\nu}\label{gamma-gamma Dirac identity} \\
&\gamma^{\mu}\gamma_{\mu}=4 \\
&\gamma^{\mu}\gamma_{5}\gamma_{\mu}=-4\gamma_{5} \\
&\gamma^{\mu}\gamma^{\nu}\gamma^{\lambda}=\eta^{\mu\nu}\gamma^{\lambda}+\eta^{\nu\lambda}\gamma^{\mu}-\eta^{\mu\lambda}\gamma^{\nu}-\rmi\epsilon^{\mu\nu\lambda\sigma}\gamma_{5}\gamma_{\sigma} \\
&\gamma^{\nu}\gamma^{\mu}\gamma_{\nu}=-2\gamma^{\mu} \\
&\gamma^{\nu}\gamma_{5}\gamma^{\mu}\gamma_{\nu}=2\gamma_{5}\gamma^{\mu} \\
&\gamma^{\mu}\gamma^{\nu}\gamma^{\sigma}\gamma^{\epsilon}=\eta^{\mu\nu}\eta^{\sigma\epsilon}+\eta^{\nu\sigma}\eta^{\mu\epsilon}-\eta^{\mu\sigma}\eta^{\nu\epsilon}-\rmi\eta^{\mu\nu}\sigma^{\sigma\epsilon}-\rmi\eta^{\nu\sigma}\sigma^{\mu\epsilon}+\rmi\eta^{\mu\sigma}\sigma^{\nu\epsilon}+\rmi\eta^{\mu\epsilon}\sigma^{\sigma\nu} \nonumber \\
&\qquad+\rmi\eta^{\nu\epsilon}\sigma^{\mu\sigma}+\rmi\eta^{\sigma\epsilon}\sigma^{\nu\mu}-\rmi\epsilon^{\mu\nu\sigma\epsilon}\gamma_{5} \\
&\gamma^{\epsilon}\sigma^{\mu\nu}=\rmi\eta^{\epsilon\mu}\gamma^{\nu}-\rmi\eta^{\epsilon\nu}\gamma^{\mu}+\epsilon^{\mu\nu\epsilon\sigma}\gamma_{5}\gamma_{\sigma} \\
&\sigma^{\mu\nu}\gamma^{\epsilon}=\rmi\eta^{\nu\epsilon}\gamma^{\mu}-\rmi\eta^{\mu\epsilon}\gamma^{\nu}+\epsilon^{\mu\nu\epsilon\sigma}\gamma_{5}\gamma_{\sigma} \\
&\gamma^{\mu}\sigma^{\sigma\epsilon}\gamma^{\nu}=\rmi\eta^{\epsilon\nu}\eta^{\mu\sigma}-\rmi\eta^{\sigma\nu}\eta^{\mu\epsilon}+\eta^{\epsilon\nu}\sigma^{\mu\sigma}-\eta^{\sigma\nu}\sigma^{\mu\epsilon}-\epsilon^{\sigma\epsilon\nu\mu}\gamma_{5}+\rmi\epsilon^{\sigma\epsilon\nu\lambda}\gamma_{5}\sigma^{\mu}{}_{\lambda} \\
&\gamma^{\sigma}\sigma^{\mu\nu}\gamma_{\sigma}=0 \\
&\sigma^{\mu\nu}\gamma_{\mu}=-3\rmi\gamma^{\nu},\label{Dirac identity sigma^mu,nu gamma_mu} \\
&\sigma^{\mu\nu}\gamma^{\rho}\gamma_{\mu}=3\rmi\eta^{\nu\rho}+\sigma^{\nu\rho},\label{Dirac identity sigma^mu,nu gamma_rho gamma_mu} \\
&\sigma^{\mu\nu}\sigma^{\rho\tau}\gamma_{\mu}=\eta^{\nu\rho}\gamma^{\tau}-\eta^{\nu\tau}\gamma^{\rho}+\rmi\epsilon^{\nu\rho\tau\sigma}\gamma_{5}\gamma_{\sigma},\label{Dirac identity sigma^mu,nu sigma^rho,tau gamma_mu} \\
&\gamma^{\mu}\sigma_{\nu\mu}=-3\rmi\gamma_{\nu},\label{Dirac identity gamma_mu sigma^nu,mu} \\
&\gamma^{\mu}\gamma^{\rho}\sigma_{\nu\mu}=3\rmi\delta_{\nu}{}^{\rho}-\sigma_{\nu}{}^{\rho},\label{Dirac identity gamma_mu gamma_rho sigma^nu,mu} \\
&\gamma^{\mu}\sigma^{\rho\tau}\sigma_{\nu\mu}=\delta_{\nu}{}^{\tau}\gamma^{\rho}-\delta_{\nu}{}^{\rho}\gamma^{\tau}+\rmi\eta_{\nu\kappa}\epsilon^{\kappa\rho\tau\sigma}\gamma_{5}\gamma_{\sigma},\label{Dirac identity gamma_mu sigma^rho,tau sigma^nu,mu} \\
&-\epsilon^{\lambda\rho\sigma\epsilon}\epsilon_{\lambda}{}^{\mu\nu\tau}=\eta^{\rho\mu}\eta^{\sigma\nu}\eta^{\epsilon\tau}-\eta^{\rho\mu}\eta^{\epsilon\nu}\eta^{\sigma\tau}+\eta^{\rho\nu}\eta^{\sigma\tau}\eta^{\epsilon\mu}-\eta^{\rho\nu}\eta^{\epsilon\tau}\eta^{\sigma\mu}+\eta^{\rho\tau}\eta^{\sigma\mu}\eta^{\epsilon\nu} \nonumber \\
&\qquad-\eta^{\rho\tau}\eta^{\epsilon\mu}\eta^{\sigma\nu}
\end{align}

\section{Derivation of the Belinfante Fierz identity}\label{Derivation of the Belinfante Fierz identity}
Here we supplement section \ref{Belinfante tensor in bilinear form} with a more detailed version of the derivation of (\ref{Belinfante Fierz identity}). The four Fierz expansions containing the term we want to solve for, $[(\partial_{\mu}\psibar)\gamma_{\nu}\psi-\psibar\gamma_{\nu}(\partial_{\mu}\psi)]$ are
\begin{align}
&j_{\nu}[\psibar(\partial_{\mu}\psi)-(\partial_{\mu}\psibar)\psi]=\frac{\rmi}{3}(\partial_{\mu}j^{\sigma})s_{\nu\sigma}-\frac{\rmi}{3}j^{\sigma}(\partial_{\mu}s_{\nu\sigma})+\frac{1}{3}(\partial_{\mu}\omega)k_{\nu}-\frac{1}{3}\omega(\partial_{\mu}k_{\nu}) \nonumber \\
&\qquad +\frac{1}{3}\sigma[\psibar\gamma_{\nu}(\partial_{\mu}\psi)-(\partial_{\mu}\psibar)\gamma_{\nu}\psi]-\frac{\rmi}{3}\sdual_{\nu\sigma}[\psibar\gamma_{5}\gamma^{\sigma}(\partial_{\mu}\psi)-(\partial_{\mu}\psibar)\gamma_{5}\gamma^{\sigma}\psi] \nonumber \\
&\qquad-\frac{\rmi}{3}k^{\sigma}[\psibar\gamma_{5}\sigma_{\nu\sigma}(\partial_{\mu}\psi)-(\partial_{\mu}\psibar)\gamma_{5}\sigma_{\nu\sigma}\psi], \\
&j_{\nu}[\psibar\gamma_{5}(\partial_{\mu}\psi)-(\partial_{\mu}\psibar)\gamma_{5}\psi]=\frac{\rmi}{3}(\partial_{\mu}j^{\sigma})\sdual_{\nu\sigma}-\frac{\rmi}{3}j^{\sigma}(\partial_{\mu}\sdual_{\nu\sigma})+\frac{1}{3}(\partial_{\mu}\sigma)k_{\nu}-\frac{1}{3}\sigma(\partial_{\mu}k_{\nu}) \nonumber \\
&\qquad+\frac{1}{3}\omega[\psibar\gamma_{\nu}(\partial_{\mu}\psi)-(\partial_{\mu}\psibar)\gamma_{\nu}\psi]-\frac{\rmi}{3}s_{\nu\sigma}[\psibar\gamma_{5}\gamma^{\sigma}(\partial_{\mu}\psi)-(\partial_{\mu}\psibar)\gamma_{5}\gamma^{\sigma}\psi] \nonumber \\
&\qquad-\frac{\rmi}{3}k^{\sigma}[\psibar\sigma_{\nu\sigma}(\partial_{\mu}\psi)-(\partial_{\mu}\psibar)\sigma_{\nu\sigma}\psi], \\
&k_{\nu}(\partial_{\mu}\sigma)=\frac{1}{3}\sigma(\partial_{\mu}k_{\nu})-\frac{\rmi}{3}\partial_{\mu}(j^{\sigma}\sdual_{\nu\sigma})+\frac{1}{3}j_{\nu}[\psibar\gamma_{5}(\partial_{\mu}\psi)-(\partial_{\mu}\psibar)\gamma_{5}\psi] \nonumber \\
&\qquad+\frac{\rmi}{3}s_{\nu\sigma}[\psibar\gamma_{5}\gamma^{\sigma}(\partial_{\mu}\psi)-(\partial_{\mu}\psibar)\gamma_{5}\gamma^{\sigma}\psi]-\frac{\rmi}{3}k^{\sigma}[\psibar\sigma_{\nu\sigma}(\partial_{\mu}\psi)-(\partial_{\mu}\psibar)\sigma_{\nu\sigma}\psi] \nonumber \\
&\qquad-\frac{1}{3}\omega[\psibar\gamma_{\nu}(\partial_{\mu}\psi)-(\partial_{\mu}\psibar)\gamma_{\nu}\psi], \\
&k_{\nu}(\partial_{\mu}\omega)=\frac{1}{3}\omega(\partial_{\mu}k_{\nu})-\frac{\rmi}{3}\partial_{\mu}(j^{\sigma}s_{\nu\sigma})+\frac{1}{3}j_{\nu}[\psibar(\partial_{\mu}\psi)-(\partial_{\mu}\psibar)\psi] \nonumber \\
&\qquad+\frac{\rmi}{3}\sdual_{\nu\sigma}[\psibar\gamma_{5}\gamma^{\sigma}(\partial_{\mu}\psi)-(\partial_{\mu}\psibar)\gamma_{5}\gamma^{\sigma}\psi]-\frac{\rmi}{3}k^{\sigma}[\psibar\gamma_{5}\sigma_{\nu\sigma}(\partial_{\mu}\psi)-(\partial_{\mu}\psibar)\gamma_{5}\sigma_{\nu\sigma}\psi] \nonumber \\
&\qquad-\frac{1}{3}\sigma[\psibar\gamma_{\nu}(\partial_{\mu}\psi)-(\partial_{\mu}\psibar)\gamma_{\nu}\psi].
\end{align}
Combining these equations gives
\begin{align}\label{Fierz identity for Belinfante, first version}
&[\psibar\gamma_{\nu}(\partial_{\mu}\psi)-(\partial_{\mu}\psibar)\gamma_{\nu}\psi]=(\sigma\omega)^{-1}\left(-\frac{\rmi}{2}(\partial_{\mu}j^{\sigma})(\omega s_{\nu\sigma}+\sigma\sdual_{\nu\sigma})-k_{\nu}[\sigma(\partial_{\mu}\sigma)+\omega(\partial_{\mu}\omega)]\right. \nonumber \\
&\qquad+\frac{1}{2}(\partial_{\mu}k_{\nu})(\sigma^{2}+\omega^{2})+j_{\nu}\{\omega[\psibar(\partial_{\mu}\psi)-(\partial_{\mu}\psibar)\psi]+\sigma[\psibar\gamma_{5}(\partial_{\mu}\psi)-(\partial_{\mu}\psibar)\gamma_{5}\psi]\} \nonumber \\
&\left.\qquad\ \ +\frac{\rmi}{2}(\sigma s_{\nu\sigma}+\omega\sdual_{\nu\sigma})[\psibar\gamma_{5}\gamma^{\sigma}(\partial_{\mu}\psi)-(\partial_{\mu}\psibar)\gamma_{5}\gamma^{\sigma}\psi]\right),
\end{align}
which obviously requires more Fierz manipulation, since there are still spinor terms present. Using the Dirac identities (\ref{Dirac identity sigma^mu,nu gamma_mu})-(\ref{Dirac identity sigma^mu,nu sigma^rho,tau gamma_mu}), we obtain the additional Fierz expansions
\begin{align}
&s_{\nu\sigma}[\psibar\gamma_{5}\gamma^{\sigma}(\partial_{\mu}\psi)-(\partial_{\mu}\psibar)\gamma_{5}\gamma^{\sigma}\psi] \nonumber \\
&\qquad=\frac{3\rmi}{5}\sigma(\partial_{\mu}k_{\nu})-\frac{3\rmi}{5}(\partial_{\mu}\sigma)k_{\nu}+\frac{1}{5}j^{\sigma}(\partial_{\mu}\sdual_{\nu\sigma})-\frac{1}{5}(\partial_{\mu}j^{\sigma})\sdual_{\nu\sigma} \nonumber \\
&\qquad\ \ +\frac{3\rmi}{5}j_{\nu}[\psibar\gamma_{5}(\partial_{\mu}\psi)-(\partial_{\mu}\psibar)\gamma_{5}\psi]-\frac{1}{5}k^{\sigma}[\psibar\sigma_{\nu\sigma}(\partial_{\mu}\psi)-(\partial_{\mu}\psibar)\sigma_{\nu\sigma}\psi] \nonumber \\
&\qquad \ \ +\frac{3\rmi}{5}\omega[\psibar\gamma_{\nu}(\partial_{\mu}\psi)-(\partial_{\mu}\psibar)\gamma_{\nu}\psi],
\end{align}
\begin{align}
&\sdual_{\nu\sigma}[\psibar\gamma_{5}\gamma^{\sigma}(\partial_{\mu}\psi)-(\partial_{\mu}\psibar)\gamma_{5}\gamma^{\sigma}\psi] \nonumber \\
&\qquad=\frac{3\rmi}{5}\omega(\partial_{\mu}k_{\nu})-\frac{3\rmi}{5}(\partial_{\mu}\omega)k_{\nu}+\frac{1}{5}j^{\sigma}(\partial_{\mu}s_{\nu\sigma})-\frac{1}{5}(\partial_{\mu}j^{\sigma})s_{\nu\sigma} \nonumber \\
&\qquad\ \ +\frac{3\rmi}{5}j_{\nu}[\psibar(\partial_{\mu}\psi)-(\partial_{\mu}\psibar)\psi]-\frac{1}{5}k^{\sigma}[\psibar\gamma_{5}\sigma_{\nu\sigma}(\partial_{\mu}\psi)-(\partial_{\mu}\psibar)\gamma_{5}\sigma_{\nu\sigma}\psi] \nonumber \\
&\qquad\ \ +\frac{3\rmi}{5}\sigma[\psibar\gamma_{\nu}(\partial_{\mu}\psi)-(\partial_{\mu}\psibar)\gamma_{\nu}\psi].
\end{align}
Combining these expansions into the form they appear in (\ref{Fierz identity for Belinfante, first version}), we get
\begin{align}\label{Rank-2 Belinfante Fierz term, requiring further Fierz expansions}
&\frac{\rmi}{2}(\sigma s_{\nu\sigma}+\omega\sdual_{\nu\sigma})[\psibar\gamma_{5}\gamma^{\sigma}(\partial_{\mu}\psi)-(\partial_{\mu}\psibar)\gamma_{5}\gamma^{\sigma}\psi]=\frac{3}{10}k_{\nu}[\sigma(\partial_{\mu}\sigma)+\omega(\partial_{\mu}\omega)] \nonumber \\
&\qquad-\frac{3}{10}(\sigma^{2}+\omega^{2})(\partial_{\mu}k_{\nu})-\frac{\rmi}{10}(\partial_{\mu}j^{\sigma})(\sigma\sdual_{\nu\sigma}+\omega s_{\nu\sigma})+\frac{\rmi}{10}j^{\sigma}[\sigma(\partial_{\mu}\sdual_{\nu\sigma})+\omega(\partial_{\mu}s_{\nu\sigma})] \nonumber \\
&\qquad+\frac{3}{10}j_{\nu}(\sigma^{2}-\omega^{2})^{-1}[j^{\sigma}(\partial_{\mu}k_{\sigma})(\sigma^{2}+\omega^{2})+2\rmi m^{\sigma}(\partial_{\mu}n_{\sigma})\sigma\omega] \nonumber \\
&\qquad-\frac{\rmi}{10}k^{\sigma}\{\sigma[\psibar\sigma_{\nu\sigma}(\partial_{\mu}\psi)-(\partial_{\mu}\psibar)\sigma_{\nu\sigma}\psi]+\omega[\psibar\gamma_{5}\sigma_{\nu\sigma}(\partial_{\mu}\psi)-(\partial_{\mu}\psibar)\gamma_{5}\sigma_{\nu\sigma}\psi]\} \nonumber \\
&\qquad-\frac{3}{5}\sigma\omega[\psibar\gamma_{\nu}(\partial_{\mu}\psi)-(\partial_{\mu}\psibar)\gamma_{\nu}\psi],
\end{align}
which itself contains terms requiring further Fierz analysis. Using the Dirac identities (\ref{Dirac identity gamma_mu sigma^nu,mu})-(\ref{Dirac identity gamma_mu sigma^rho,tau sigma^nu,mu}), we find that the expansion of these terms is
\begin{align}
&k^{\sigma}[\psibar\sigma_{\nu\sigma}(\partial_{\mu}\psi)-(\partial_{\mu}\psibar)\sigma_{\nu\sigma}\psi]\!=\!\frac{1}{5}(\partial_{\mu}j^{\sigma})\sdual_{\nu\sigma}-\frac{1}{5}j^{\sigma}(\partial_{\mu}\sdual_{\nu\sigma})+\frac{3\rmi}{5}(\partial_{\mu}\sigma)k_{\nu} \nonumber \\
&\qquad-\frac{3\rmi}{5}\sigma(\partial_{\mu}k_{\nu})+\frac{3\rmi}{5}j_{\nu}[\psibar\gamma_{5}(\partial_{\mu}\psi)-(\partial_{\mu}\psibar)\gamma_{5}\psi]-\frac{1}{5}s_{\nu\sigma}[\psibar\gamma_{5}\gamma^{\sigma}(\partial_{\mu}\psi)-(\partial_{\mu}\psibar)\gamma_{5}\gamma^{\sigma}\psi] \nonumber \\
&\qquad+\frac{3\rmi}{5}\omega[\psibar\gamma_{\nu}(\partial_{\mu}\psi)-(\partial_{\mu}\psibar)\gamma_{\nu}\psi], \\
&k^{\sigma}[\psibar\gamma_{5}\sigma_{\nu\sigma}(\partial_{\mu}\psi)-(\partial_{\mu}\psibar)\gamma_{5}\sigma_{\nu\sigma}\psi]=\frac{1}{5}(\partial_{\mu}j^{\sigma})s_{\nu\sigma}-\frac{1}{5}j^{\sigma}(\partial_{\mu}s_{\nu\sigma})+\frac{3\rmi}{5}(\partial_{\mu}\omega)k_{\nu} \nonumber \\
&\qquad-\frac{3\rmi}{5}\omega(\partial_{\mu}k_{\nu})+\frac{3\rmi}{5}j_{\nu}[\psibar(\partial_{\mu}\psi)-(\partial_{\mu}\psibar)\psi]-\frac{1}{5}\sdual_{\nu\sigma}[\psibar\gamma_{5}\gamma^{\sigma}(\partial_{\mu}\psi)-(\partial_{\mu}\psibar)\gamma_{5}\gamma^{\sigma}\psi] \nonumber \\
&\qquad+\frac{3\rmi}{5}\sigma[\psibar\gamma_{\nu}(\partial_{\mu}\psi)-(\partial_{\mu}\psibar)\gamma_{\nu}\psi].
\end{align}
Again, combining these terms into the form in which they appear in (\ref{Rank-2 Belinfante Fierz term, requiring further Fierz expansions}) gives
\begin{align}
&-\frac{\rmi}{10}k^{\sigma}\{\sigma[\psibar\sigma_{\nu\sigma}(\partial_{\mu}\psi)-(\partial_{\mu}\psibar)\sigma_{\nu\sigma}\psi]+\omega[\psibar\gamma_{5}\sigma_{\nu\sigma}(\partial_{\mu}\psi)-(\partial_{\mu}\psibar)\gamma_{5}\sigma_{\nu\sigma}\psi]\} \nonumber \\
&\ =\frac{\rmi}{50}(\sigma s_{\nu\sigma}+\omega\sdual_{\nu\sigma})[\psibar\gamma_{5}\gamma^{\sigma}(\partial_{\mu}\psi)-(\partial_{\mu}\psibar)\gamma_{5}\gamma^{\sigma}\psi]+\frac{6}{50}\sigma\omega[\psibar\gamma_{\nu}(\partial_{\mu}\psi)-(\partial_{\mu}\psibar)\gamma_{\nu}\psi] \nonumber \\
&\qquad-\frac{3}{50}j_{\nu}(\sigma^{2}-\omega^{2})^{-1}[j^{\sigma}(\partial_{\mu}k_{\sigma})(\sigma^{2}+\omega^{2})+2\rmi m^{\sigma}(\partial_{\mu}n_{\sigma})\sigma\omega]-\frac{3}{50}(\sigma^{2}+\omega^{2})(\partial_{\mu}k_{\nu}) \nonumber \\
&\qquad+\frac{3}{50}k_{\nu}[\sigma(\partial_{\mu}\sigma)+\omega(\partial_{\mu}\omega)]+\frac{\rmi}{50}j^{\sigma}[\sigma(\partial_{\mu}\sdual_{\nu\sigma})+\omega(\partial_{\mu}s_{\nu\sigma})] \nonumber \\
&\qquad-\frac{\rmi}{50}(\partial_{\mu}j^{\sigma})(\sigma\sdual_{\nu\sigma}+\omega s_{\nu\sigma}),
\end{align}
which when substituting into (\ref{Rank-2 Belinfante Fierz term, requiring further Fierz expansions}) and rearranging, gives
\begin{align}\label{Rank-2 bilinear and spinor dependent Fierz identity}
&\frac{\rmi}{2}(\sigma s_{\nu\sigma}+\omega\sdual_{\nu\sigma})[\psibar\gamma_{5}\gamma^{\sigma}(\partial_{\mu}\psi)-(\partial_{\mu}\psibar)\gamma_{5}\gamma^{\sigma}\psi]=-\frac{1}{2}\sigma\omega[\psibar\gamma_{\nu}(\partial_{\mu}\psi)-(\partial_{\mu}\psibar)\gamma_{\nu}\psi] \nonumber \\
&\ +\frac{1}{4}j_{\nu}(\sigma^{2}-\omega^{2})^{-1}[j^{\sigma}(\partial_{\mu}k_{\sigma})(\sigma^{2}+\omega^{2})+2\rmi m^{\sigma}(\partial_{\mu}n_{\sigma})\sigma\omega]-\frac{3}{8}(\sigma^{2}+\omega^{2})(\partial_{\mu}k_{\nu}) \nonumber \\
&\ +\frac{3}{8}k_{\nu}[\sigma(\partial_{\mu}\sigma)+\omega(\partial_{\mu}\omega)]-\frac{\rmi}{8}(\partial_{\mu}j^{\sigma})(\sigma\sdual_{\nu\sigma}+\omega s_{\nu\sigma})+\frac{\rmi}{8}j^{\sigma}[\sigma(\partial_{\mu}\sdual_{\nu\sigma})+\omega(\partial_{\mu}s_{\nu\sigma})],
\end{align}
a pure bilinear tensor expression. Now, using the Fierz identities derived in \cite{InglisJarvis2014}
\begin{align}
&[\psibar(\partial_{\mu}\psi)-(\partial_{\mu}\psibar)\psi]=-(\sigma^{2}-\omega^{2})^{-1}[j^{\nu}(\partial_{\mu}k_{\nu})\omega+\rmi m^{\nu}(\partial_{\mu}n_{\nu})\sigma], \\
&[\psibar\gamma_{5}(\partial_{\mu}\psi)-(\partial_{\mu}\psibar)\gamma_{5}\psi]=-(\sigma^{2}-\omega^{2})^{-1}[j^{\nu}(\partial_{\mu}k_{\nu})\sigma+\rmi m^{\nu}(\partial_{\mu}n_{\nu})\omega],
\end{align}
and combining them into the form in which they appear in (\ref{Fierz identity for Belinfante, first version}), we get
\begin{align}
&\omega[\psibar(\partial_{\mu}\psi)-(\partial_{\mu}\psibar)\psi]+\sigma[\psibar\gamma_{5}(\partial_{\mu}\psi)-(\partial_{\mu}\psibar)\gamma_{5}\psi] \nonumber \\
&\qquad=-(\sigma^{2}-\omega^{2})^{-1}[j^{\sigma}(\partial_{\mu}k_{\sigma})(\sigma^{2}+\omega^{2})+2\rmi m^{\sigma}(\partial_{\mu}n_{\sigma})\sigma\omega],
\end{align}
which along with (\ref{Rank-2 bilinear and spinor dependent Fierz identity}), can be substituted into (\ref{Fierz identity for Belinfante, first version}) to give
\begin{align}\label{Fierz identity for Belinfante, second version}
&[\psibar\gamma_{\nu}(\partial_{\mu}\psi)-(\partial_{\mu}\psibar)\gamma_{\nu}\psi] \nonumber \\
&\qquad=(\sigma\omega)^{-1}\left\{-\frac{1}{2}j_{\nu}(\sigma^{2}-\omega^{2})^{-1}[j^{\sigma}(\partial_{\mu}k_{\sigma})(\sigma^{2}+\omega^{2})+2\rmi m^{\sigma}(\partial_{\mu}n_{\sigma})\sigma\omega]\right. \nonumber \\
&\qquad\qquad+\frac{1}{12}(\sigma^{2}+\omega^{2})(\partial_{\mu}k_{\nu})-\frac{5}{12}k_{\nu}[\sigma(\partial_{\mu}\sigma)+\omega(\partial_{\mu}\omega)] \nonumber \\
&\left.{}\qquad\qquad+\frac{\rmi}{12}j^{\sigma}[\sigma(\partial_{\mu}\sdual_{\nu\sigma})+\omega(\partial_{\mu}s_{\nu\sigma})]-\frac{5\rmi}{12}(\partial_{\mu}j^{\sigma})(\sigma\sdual_{\nu\sigma}+\omega s_{\nu\sigma})\right\}.
\end{align}
This expression contains no explicit spinor terms, as required, but we can improve it by eliminating the rank-2 tensors $s_{\mu\nu}$ and $\sdual_{\mu\nu}$, by using the Fierz identity
\begin{align}
&s^{\mu\nu}=(\sigma^{2}-\omega^{2})^{-1}(\sigma\epsilon^{\mu\nu\rho\sigma}-\omega\delta^{\mu\nu\rho\sigma})j_{\rho}k_{\sigma}, \\
&\sdual^{\mu\nu}=(\sigma^{2}-\omega^{2})^{-1}(\omega\epsilon^{\mu\nu\rho\sigma}-\sigma\delta^{\mu\nu\rho\sigma})j_{\rho}k_{\sigma},
\end{align}
where we define the partially antisymmetric object
\begin{equation}
\delta^{\mu\nu\rho\sigma}\equiv\rmi(\eta^{\mu\rho}\eta^{\nu\sigma}-\eta^{\mu\sigma}\eta^{\nu\rho}).
\end{equation}
We also require the derivatives of these identities, which are
\begin{align}
\partial_{\mu}s_{\nu\sigma}={}&(\sigma^{2}-\omega^{2})^{-2}\{[2\sigma\omega(\partial_{\mu}\omega)-(\sigma^{2}+\omega^{2})(\partial_{\mu}\sigma)]\epsilon_{\nu\sigma\rho\epsilon} \nonumber \\
&+[2\sigma\omega(\partial_{\mu}\sigma)-(\sigma^{2}+\omega^{2})(\partial_{\mu}\omega)]\delta_{\nu\sigma\rho\epsilon}\}j^{\rho}k^{\epsilon} \nonumber \\
&+(\sigma^{2}-\omega^{2})^{-1}(\sigma\epsilon_{\nu\sigma\rho\epsilon}-\omega\delta_{\nu\sigma\rho\epsilon})[(\partial_{\mu}j^{\rho})k^{\epsilon}+j^{\rho}(\partial_{\mu}k^{\epsilon})], \\
\partial_{\mu}\sdual_{\nu\sigma}={}&(\sigma^{2}-\omega^{2})^{-2}\{[-2\sigma\omega(\partial_{\mu}\sigma)+(\sigma^{2}+\omega^{2})(\partial_{\mu}\omega)]\epsilon_{\nu\sigma\rho\epsilon} \nonumber \\
&+[-2\sigma\omega(\partial_{\mu}\omega)+(\sigma^{2}+\omega^{2})(\partial_{\mu}\sigma)]\delta_{\nu\sigma\rho\epsilon}\}j^{\rho}k^{\epsilon} \nonumber \\
&+(\sigma^{2}-\omega^{2})^{-1}(\omega\epsilon_{\nu\sigma\rho\epsilon}-\sigma\delta_{\nu\sigma\rho\epsilon})[(\partial_{\mu}j^{\rho})k^{\epsilon}+j^{\rho}(\partial_{\mu}k^{\epsilon})].
\end{align}
the rank-2 dependent terms in (\ref{Fierz identity for Belinfante, second version}) become
\begin{align}
&\frac{\rmi}{12}j^{\sigma}[\sigma(\partial_{\mu}\sdual_{\nu\sigma})+\omega(\partial_{\mu}s_{\nu\sigma})]=\frac{1}{12}(\sigma^{2}-\omega^{2})^{-1}\left\{2\sigma\omega k_{\nu}[\omega(\partial_{\mu}\sigma)-\sigma(\partial_{\mu}\omega)]\right. \nonumber \\
&\left.{}\qquad+2\rmi\sigma\omega\epsilon_{\nu\sigma\rho\epsilon}j^{\sigma}(\partial_{\mu}j^{\rho})k^{\epsilon}+j_{\nu}j^{\sigma}(\partial_{\mu}k_{\sigma})(\sigma^{2}+\omega^{2})\right\} -\frac{1}{12}(\partial_{\mu}k_{\nu})(\sigma^{2}+\omega^{2}), \\
&-\frac{5\rmi}{12}(\partial_{\mu}j^{\sigma})(\sigma\sdual_{\nu\sigma}+\omega s_{\nu\sigma})=(\sigma^{2}-\omega^{2})^{-1}\left\{\frac{5}{12}j_{\nu}j^{\sigma}(\partial_{\mu}k_{\sigma})(\sigma^{2}+\omega^{2})\right. \nonumber \\
&\left.\qquad-\frac{5\rmi}{6}\sigma\omega\epsilon_{\nu\sigma\rho\epsilon}(\partial_{\mu}j^{\sigma})j^{\rho}k^{\epsilon}-\frac{5}{12}k_{\nu}(\sigma^{2}+\omega^{2})[\omega(\partial_{\mu}\omega)-\sigma(\partial_{\mu}\sigma)]\right\},
\end{align}
giving us the final form of our identity
\begin{align}
&[\psibar\gamma_{\nu}(\partial_{\mu}\psi)-(\partial_{\mu}\psibar)\gamma_{\nu}\psi]=(\sigma^{2}-\omega^{2})^{-1}\{k_{\nu}[\omega(\partial_{\mu}\sigma)-\sigma(\partial_{\mu}\omega)]-\rmi\epsilon_{\nu\sigma\rho\epsilon}(\partial_{\mu}j^{\sigma})j^{\rho}k^{\epsilon} \nonumber \\
&\qquad-\rmi j_{\nu}m^{\sigma}(\partial_{\mu}n_{\sigma})\}.
\end{align}

\bibliographystyle{plain}
\bibliography{MD_bib}

\end{document}